\documentclass[11pt,a4paper]{article}
\pdfoutput=1

\usepackage{jheppub}
\usepackage{subfigure}
\usepackage{cancel}
\usepackage{mathtools}

\newcommand\be{\begin{equation}}
\newcommand\bea{\begin{eqnarray}}
\newcommand\ee{\end{equation}}
\newcommand\eea{\end{eqnarray}}
\newcommand{\nn}{\nonumber \\}
\newcommand{\f}[2]{\frac{#1}{#2}}
\newcommand{\bref}[1]{(\ref{#1})}
\newcommand\h{\frac{1}{2}}

\newcommand{\trho}{\tilde \rho}
\newcommand{\ttheta}{\tilde \theta}
\newcommand{\tphi}{\tilde \phi}

\newcommand{\rr}{{r_{\text{\tiny R}}}}
\newcommand{\tr}{{t_{\text{\tiny R}}}}
\newcommand{\xr}{{x_{\text{\tiny R}}}}
\newcommand{\rp}{{r_{\text{\tiny P}}}}
\newcommand{\tp}{{t_{\text{\tiny P}}}}
\newcommand{\tpe}{{t_{\text{{\tiny P,E}}}}}
\newcommand{\xp}{{x_{\text{\tiny P}}}}
\newcommand{\rsr}{{r_{\text{\tiny SR}}}}
\newcommand{\tsr}{t_{{\text{\tiny SR}}}}

\title{When UV and IR Collide:\\ Inequivalent CFTs From Different Foliations Of AdS}
\author[a]{Borun D. Chowdhury}
\author[a,b]{and Maulik K. Parikh}
\affiliation[a]{Department of Physics \\
Arizona State University \\
Tempe, Arizona 85287, USA
}
\affiliation[b]{Beyond Center for Fundamental Concepts in Science \\
Arizona State University \\
Tempe, Arizona 85287, USA
}

\abstract{In the AdS/CFT correspondence, CFTs are identified by asymptotic boundary surfaces and the boundary conditions imposed on those surfaces. However, AdS can be foliated in various ways to give different boundaries. We show that the CFTs obtained using certain distinct foliations are different. This difference arises because the asymptotic region of a foliation overlaps with the deep interior region of another. In particular we focus on the CFTs defined on surfaces of large constant radius in global coordinates, Rindler-AdS coordinates, and Poincar\'e coordinates for AdS$_3$. We refer to these as global-CFT, Rindler-CFT and Poincar\'e-CFT respectively. We demonstrate that the correlators for these CFTs are different and argue that the bulk duals to these should agree up to very close to the respective horizons but then start differing. Since the BTZ black hole is obtained as a quotient of AdS$_3$, we discuss the implications of our results for bulk duals of periodically-identified Poincar\'e and Rindler-CFTs. Our results are consistent with some recent proposals suggesting a modification of the semi-classical BTZ geometry close to the horizons.
}

\emailAdd{bdchowdh@asu.edu}
\emailAdd{maulik.parikh@asu.edu}


\begin{document}
\maketitle

\section{Introduction}

In Lorentzian AdS/CFT, the definition of a dual CFT involves  the  conformal boundary surface on which it is supposed to ``live''~\cite{Balasubramanian:1999re} and  the boundary conditions on that surface~\cite{Balasubramanian:1998sn,Compere:2008us,Compere:2013bya,Avery:2013dja}. However, different foliations of AdS result in different boundaries~\cite{Emparan:1999pm,Chowdhury:2014csa}. For example one definition of a  boundary is on a surface of large radius in global coordinates, another is  on a surface of large radius in Poincar\'e coordinates and yet another  is on a surface of large radius in Rindler-AdS coordinates.  We demonstrate that the CFTs on these surfaces are different even when the surfaces are taken to infinity (as is required to define a CFT). We refer to these CFTs as the global-CFT, the Poincar\'e-CFT and the Rindler-CFT respectively. 
For simplicity we  discuss only AdS$_3$ which is also easiest to visualise, but many of the results are generalisable to higher dimensions.

\begin{figure}[htbp] %  figure placement: here, top, bottom, or page
   \centering
   \includegraphics[width=2in]{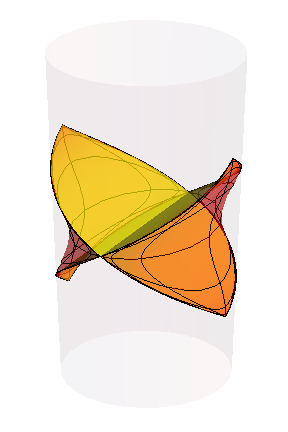}
       \caption{The boundary of AdS for global foliation is shown in gray. This is a surface of constant radius in global coordinates for large value of the radius. The yellow surfaces are surfaces of constant Rindler-AdS radius. It is easy to see that fixing boundary conditions on the global boundary imposes conditions on small Rindler-AdS radial surfaces also. Conversely, putting boundary conditions on just large Rindler-AdS radial surfaces does not put any boundary conditions on the global boundary outside a finite domain. Thus the global-CFT and the Rindler-CFT are different and imply different dynamics for the bulk.} 
   \label{fig:Rindler-Global-Intro}
\end{figure}

There is a quick way to see that the aforementioned CFTs are different (we give more details later). A CFT is defined on a constant radial surface by performing a Fefferman-Graham~\cite{FeffermanGraham} expansion for some foliation, keeping the leading term fixed and letting the subleading terms fluctuate~\cite{Balasubramanian:1999re}. It turns out that every large global radius surface invariably intersects surfaces of arbitrarily small Rindler-AdS radius~\cite{Chowdhury:2014csa} (see figure~\ref{fig:Rindler-Global-Intro}). 
Thus, when defining the global-CFT one invariably ends up imposing conditions on subleading terms in the Fefferman-Graham expansion for the Rindler-AdS foliation. Moreover, in some ranges of parameters, the Fefferman-Graham expansion itself breaks down due to small Rindler-AdS radius. Conversely, defining the Rindler-CFT one does not impose any conditions on the global boundary outside a finite domain. Similar arguments hold for the Poincar\'e-CFT.

Bulk horizons projected onto the global boundary give the edges of the so-called causal diamonds. In section~\ref{Sec:DiffBoundDiffCFT}, we demonstrate the mismatch of Fefferman-Graham expansions between various foliations at the edges of the causal diamonds. The width of the mismatch region is controlled by the UV cutoff. We also demonstrate how in the vicinity of the center of the causal diamonds, the various CFTs can be viewed as conformally related. However, it should be noted that due to the incompatibility of Fefferman-Graham expansions at the edges of the causal diamond, the various CFTs cannot globally be related by conformal transformations and are thus not truly equivalent.

In section~\ref{Sec:DifferencesInCFTs} we restrict the global CFT to be within the causal diamond by focusing on causal developments of subregions; we refer to these as Rindlerized-global-CFTs and Poincarized-global-CFT. Correlators within these can be analytically continued to the entire global boundary cylinder.  We demonstrate how the correlators of Poincarized-global-CFT (Rindlerized-global-CFTs) are approximately the same as those of Poincar\'e-CFT (Rindler-CFTs) deep inside the causal diamonds but differ at the edges. We regard this as evidence that, generically, the correlators of Poincar\'e-CFT (Rindler-CFT) cannot be analytically continued to outside the causal diamonds.

In section~\ref{sec:RelatingCFTs} we expound on how the CFTs are different. We argue that seen as part of the global-CFT there is an interaction between the Hilbert spaces associated with the Rindlerized-global-CFTs  whereas the Hilbert spaces associated with the Rindler-CFTs are not interacting. Thus, it may be possible to view the Rindler-CFTs as a deformation of the global-CFT which breaks the concerned interaction. Similar ideas apply for the Poincar\'e-CFT. We hope to come back to this issue in the future.

In section~\ref{Sec:ConjectureForBulk} we investigate the implications for the bulk physics. Since the correlators of these CFTs are different, the bulk duals to these CFTs should be different also. As the CFTs differ at the edges of the causal diamonds, and since the causal diamonds are projections of bulk horizons onto the global boundary, we conjecture that the bulk dual of Rindler-CFT and Poincar\'e-CFT should resemble semi-classical global AdS till very close to the respective horizons and then start differing. The width of the transition region is governed by the UV cutoff.

This result is particularly interesting in the context of AdS$_3$, where it becomes relevant to black holes. The BTZ black hole~\cite{Banados:1992wn} can be viewed as a quotient of AdS$_3$ space~\cite{Banados:1992gq}. The massless and massive ones come from foliating in Poincar\'e and Rindler-AdS coordinates respectively and periodically identifying along a spatial isometry. The massless BTZ black hole has a singular horizon because of vanishing size so the region behind the horizon is not accessible in supergravity. However, for the massive BTZ black hole the identification produces orbifold singularities (interpreted as the eternal black hole singularity) behind the horizons  (interpreted as the eternal black hole event horizons)  but is innocuous on the horizons themselves. Thus, one may be inclined to think that dynamics involving horizon-crossing in global AdS$_3$ might carry over trivially to BTZ. One may further be inclined to think that such dynamics and more generally the interior of the BTZ may be captured by quotients of global-CFT~\cite{Horowitz:1998xk,KeskiVakkuri:1998nw,Maldacena:2001kr,Hemming:2002kd}.

However, the discrete symmetry relevant for orbifolding to obtain the BTZ black holes are isometries of constant Poincar\'e and Rindler-AdS radial surfaces and not of surfaces of constant global radius. Thus a natural question is:  what are the bulk duals to periodically identified Poincar\'e-CFT (PIPC) and periodically identified Rindler-CFTs (PIRCs)? The dynamics of the quotiented bulk duals to global-CFT (i.e. the BTZ black holes) cannot be trivially assumed to give the dynamics of the bulk duals to PIPC and PIRCs.  After periodic identification the edges of the causal diamonds correspond to large times so our results indicate that the correlators of PIPC and PIRCs differ from the naive ones found from the BTZ geometries at late times~\cite{KeskiVakkuri:1998nw}. Furthermore, our conjecture implies that bulk duals to these CFTs will resemble the massless and massive BTZ respectively till very close to the horizon and then start differing. 

In section~\ref{sec:MasslessBTZ} we focus on the massless BTZ black hole. In the case of the D1-D5 system the near-horizon naive geometry is the massless BTZ $\times S^3 \times T^4$ and the actual geometries are the Lunin-Mathur geometries. The typical ones resemble the massless BTZ black hole till very close to the ``would-be'' horizon and then start differing.  In addition, the PIPC correlators dual to these geometries  show the late-time deviations from the naive ones. So in hindsight, the ideas stated in the previous paragraph have already been realised for this case.

In section~\ref{sec:MassiveBTZ} we focus on the massive BTZ black hole. The story for the massive BTZ black hole is not as well settled as for the massless case. Recently Ref.~\cite{Kay:2013gia} has claimed that the bulk is unstable to small fluctuations. Further, problems related to the holographic relation between the bulk and boundary proposed in~\cite{Maldacena:2001kr} (see also~\cite{Balasubramanian:1998de}) have been raised in~\cite{Marolf:2012xe,Avery:2013bea,Mathur:2014dia}. In fact,~\cite{Kay:2013gia,Avery:2013bea,Mathur:2014dia} have made conjectures which amount to claiming that the dual to PIRCs have the regions behind the horizons removed and end in capped (quantum) geometries beyond the would-be horizons (see also~\cite{Hawking:2014tga,Kabat:2014kfa}).  Our results are consistent with the proposals of~\cite{Kay:2013gia,Avery:2013bea,Mathur:2014dia} and raise further issues with the proposal of~\cite{Maldacena:2001kr}.

\section{Different boundaries and different CFTs} \label{Sec:DiffBoundDiffCFT}

\subsection{The boundary CFT} \label{Sec:DefBoundary} 

To equate the dynamics in AdS to those in a CFT one needs the so-called dictionary between them. The first entry in this dictionary is the definition of the ``boundary'' on which the CFT is supposed to live (loosely speaking, since AdS and CFT are dual descriptions). Asymptotically AdS spacetimes admit a Fefferman-Graham expansion of their metrics:
\be
ds^2 \rightarrow \f{dr^2}{r^2} + \left(r^2 g^{(0)}_{ab} + g^{(2)}_{ab} + \mathcal O(r^{-2}) \right) d x^a d x^b . \label{FeffermanGraham}
\ee
The boundary is understood to be at a fixed large value of $r$ that we refer to as $r_c$. This location is related to the cutoff of the dual theory and a CFT is obtained by taking $r_c \to \infty$. The coordinates $x^a$ span the field theory directions. Since the metric blows up for large $r$ the metric on AdS does not define a metric on the boundary but instead yields a conformal structure. Thus, $g^{(0)}_{ab}$ is the boundary metric up to Weyl transformations~\cite{Witten:1998qj}. The on-shell variation of the gravity action, which includes the Einstein-Hilbert term, the Gibbons-Hawking term, and a divergence-cancelling counter-term~\cite{Balasubramanian:1999re},
\be
S=\f{1}{16 \pi G} \int d^{d+1} x ~ \sqrt{g^{(d+1)}} (R-2 \Lambda)  + \f{1}{8\pi G} \int_{\partial M} d^d x ~ \sqrt{g^{(d)}} K + \f{1}{8 \pi G} S_{ct} (g^{(d)}) ~,
\ee
gives
\be
\delta S = \f{1}{2} \int_{\partial M}  d^2 x \sqrt{-g^{(0)}} T^{ab} \delta g^{(0)}_{ab} ~,
\ee
where $T^{ab}$ is a symmetric tensor that is interpreted as the expectation value of the stress tensor of the CFT~\cite{Balasubramanian:1999re}. The variational principle is well-posed if we impose Dirichlet boundary conditions $\delta g^{(0)}_{ab}=0$.\footnote{Certain other boundary conditions are also allowed~\cite{Compere:2008us,Compere:2013bya,Avery:2013dja} but for simplicity we only discuss Dirichlet boundary conditions.} 

Imposing a boundary condition specifies the theory; Dirichlet boundary conditions in particular amount to ``holding the boundary fixed''~\cite{Compere:2013bya}. $g^{(2)}_{ab}$ is allowed to fluctuate and captures information of the state. 
In fact for flat boundaries, $T^{ab} \sim g^{(2)}_{ab}$ and this information is thus encoded in the stress tensor.

Implicit in the choice of boundary conditions is the choice of surface on which such conditions are being imposed. While \bref{FeffermanGraham} does not allow such a choice, that is because the choice has already been made by foliating spacetime in a particular way. We will discuss more about this issue of choice of boundary surfaces below.

\subsection{States in the boundary CFT} \label{Sec:BrownHenneaux} 

In the special case of AdS$_3$ we have another way to understand the boundary conditions. Brown and Henneaux~\cite{Brown:1986nw} have shown that diffeomorphisms with the asymptotic (large $r$) form:
\bea
x^+ \to x^+ - \xi^+ -\f{1}{2r^2} \partial^2_-\xi^- ~,\label{BH1} \\
x^- \to x^- - \xi^- -\f{1}{2r^2} \partial^2_+ \xi^+  ~, \label{BH2} \\
r \to r + \f{r}{2} (\partial_+ \xi^+ + \partial_- \xi^-) ~,  \label{BH3}
\eea
where $x^\pm = t \pm x$, preserve the asymptotic boundary conditions:
\begin{gather}
g_{+-} = - \f{r^2}{2} +\mathcal O(1)~, \qquad g_{++}=\mathcal O(1)~, \qquad g_{--} =\mathcal O(1)~, \nn
g_{rr} = \f{1}{r^2} + \mathcal O(r^{-4})~, \qquad g_{+r} = \mathcal O(r^{-3})~,  \qquad g_{-r} = \mathcal O(r^{-3})~.
\end{gather}
In the $r \to \infty$ limit the transformations of $x^\pm$ induce conformal transformations on the boundary CFT and this is reflected in changes in $g^{(2)}$ while at the same time keeping $\delta g^{(0)}=0$.

We see from~\bref{BH3} that conformal transformations inducing diffeomorphisms change the location of the boundary surface. So different foliations permitting asymptotic Fefferman-Graham forms do not immediately imply that the associated CFTs are genuinely different. In particular they are not different if they are related by a Brown-Henneaux transformation as then there is a conformal mapping between the two.\footnote{We thank Nemani Suryanarayana for discussions on this point.}

%%% ---

\subsection{Global vs. Rindler boundary} 

\subsubsection{Global and Rindler-AdS foliations and their boundaries}

One can write the global AdS$_3$ line element as\footnote{For definiteness we discuss only three-dimensional AdS. The results can be generalised to higher dimensions in a straightforward way for most of the paper. An exception is the discussion of the BTZ black hole which can be viewed as a quotient of AdS$_3$; higher-dimensional eternal AdS black holes cannot be obtained as quotients of AdS.}
\be
ds^2= \f{d \rho^2}{\rho^2+1} -(\rho^2+1) d\tau^2 + \rho^2 d \phi^2 \label{globalAdS}
\ee
where $\rho \in [0,\infty)$, $\tau \in (-\infty,\infty)$ and $\phi \sim \phi + 2\pi.$\footnote{All throughout this paper we set the AdS radius to unity.} These coordinates cover the entire manifold. It is often useful to conformally compactify $\rho$ and visualise AdS$_3$  as a solid cylinder (figure~\ref{fig:Rindler-Global-Intro}). Similarly, one can also write the metric for the Rindler-AdS wedges in BTZ form:
\be
ds^2 =  \f{d \rr^2}{\rr^2-1} - (\rr^2-1)d\tr^2 + \rr^2 d \xr^2 \label{RindlerAdS}
\ee
where $\rr \in (1,\infty)$, $\tr \in (-\infty,\infty)$ and $\xr \in (-\infty,\infty)$. There is an acceleration horizon at $\rr=1$ and these coordinates cover the region outside the horizon~\cite{Emparan:1999pm,Emparan:1999gf,Hamilton:2006az,Myers:2010xs,Parikh:2012kg}. The rest of AdS$_3$ may be viewed as a Kruskal-like extension of these coordinates. The temperature associated with the acceleration horizon can be read off by Wick rotation and demanding the absence of a conical singularity; it turns out to be $\f{1}{2\pi}$.\footnote{There are ways to foliate AdS in ways which give an inner horizon as well~\cite{Hemming:2002kd,Parikh:2011aa} that corresponds to rotating BTZ string. For simplicity we will not discuss those.}

One can perform a large $\rho$ expansion to write  \bref{globalAdS} in the Fefferman-Graham form~\cite{FeffermanGraham} and define a CFT on the cylindrical boundary $S^1 \times R$ at $\rho=\rho_c \to \infty$. One can also perform a large $\rr$ expansion to write \bref{RindlerAdS} in the Fefferman-Graham form and define two CFTs on $R^{1,1} \times R^{1,1}$ at $\rr=\rr_c \to \infty$.\footnote{\label{fn:changeRadialCood}Care must be taken to redefine the radial coordinates to put the metric in the Fefferman-Graham form to read off the values of $g^{(0)}_{ab}$ and $g^{(2)}_{ab}$.} We refer to the former CFT as the global-CFT and the latter CFT pair as the Rindler-CFTs. 

It has been claimed that these two CFTs are equivalent  (see~\cite{Balasubramanian:1998sn,Barbon:2013nta} for example). We will argue that this is not the case.

\begin{figure}[htbp] %  figure placement: here, top, bottom, or page
   \centering
   \subfigure[Finite cutoff surfaces]{
   \includegraphics[width=2in]{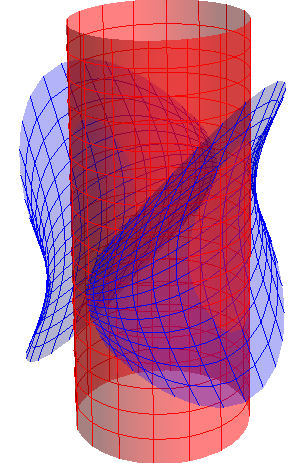}} 
   \hspace{1in}
   \subfigure[Infinite cutoff surfaces]{
   \includegraphics[width=1.6in]{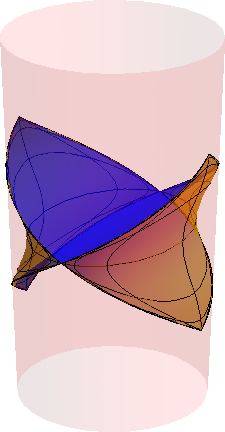}}
    \caption{(a) Global cutoff surface $\rho=\rho_c$ is shown in red and Rindler cutoff surface $\rr=\rr_c$ is shown in blue. (b) When we take $\rho_c$ to infinity, all the $\rr$ surfaces bunch up along the edges of the ``causal diamond". Two of them are shown in the figure.}
   \label{fig:CutoffSurfaces}
\end{figure}

Let us begin by writing down the expressions relating global to Rindler-AdS coordinates:
\bea
\rho^2 &=& (\rr^2-1) \f{\cosh 2 \xr + \cosh 2 \tr}{2} + \sinh^2 \xr ~, \label{RindlerToGlobal1}\\
\cot \phi &=&- \f{\rr}{\sqrt{\rr^2-1}} \f{\sinh \xr  }{\cosh \tr}  ~,\label{RindlerToGlobal2} \\
\tan \tau &=&\f{\sqrt{\rr^2-1}}{\rr} \f{\sinh \tr  }{\cosh \xr} ~,\label{RindlerToGlobal3}
\eea
and the inverse relations:
\bea
\rr^2-1&=& - (\rho^2+1) \sin^2 \tau  + \rho^2 \sin^2 \phi   ~, \label{GlobalToRindler1} \\
\tanh \xr &=&- \f{\rho}{\sqrt{\rho^2+1}} \f{\cos \phi}{\cos \tau}  ~, \label{GlobalToRindler2} \\
\tanh \tr &=& \f{\sqrt{\rho^2+1}}{\rho} \f{\sin \tau}{\sin \phi}  ~. \label{GlobalToRindler3}
\eea
From~\bref{GlobalToRindler1} it is clear that for any given $\rho_c$, one gets  $\rr=1$ (the Rindler-AdS horizon) for suitable values of $\phi$ and $\tau$.
Said differently, the bulk acceleration horizon intersects the  cylinder of any radius $\rho_c$ and imposing boundary conditions on the cylinder to define the global-CFT will always impose conditions on the metric for small $\rr$ when $\xr,\tr$ are large enough. This behaviour persists when  $\rho_c \to \infty$. This is shown in figure~\ref{fig:CutoffSurfaces}b. If on the other hand one wants to define the Rindler-CFT then one needs to take large $\rr_c$ and permit arbitrarily large $\xr$ and $\tr$. Then one takes $\rr_c \to \infty$. This is clearly not consistent with the above procedure. This justifies our claim that the Rindler-CFT pair is different from the global-CFT.

\subsubsection{Global-CFT vs. Rindler-CFT}

\begin{figure}[htbp] %  figure placement: here, top, bottom, or page
   \centering
   \includegraphics[width=3.5in]{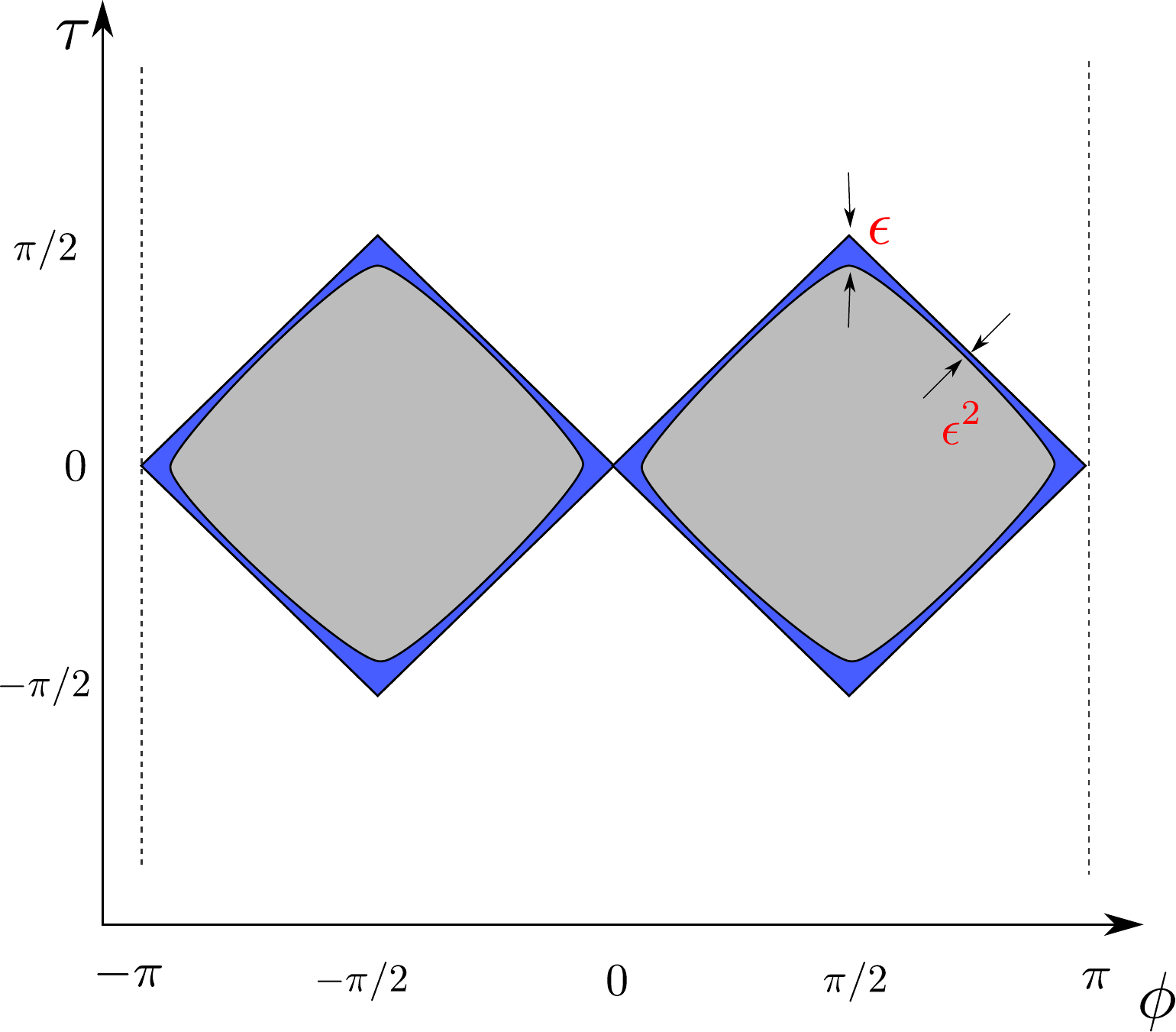}   
    \caption{We open up the global boundary cylinder for better visualisation. The coloured regions are the causal diamonds and are the interior of the curve the bulk Rindler-AdS horizons trace on the boundary cylinder. The global boundary cylinder is taken to be large with $\rho \sim \mathcal O(\epsilon^{-1})$ and $\epsilon \ll 1$. The regions where the Rindler-AdS radial coordinate $\rr \sim \mathcal O(1)$ is shown in blue. This is the region where the Fefferman-Graham expansion in $\rho$ and $\rr$ are not consistent and the Rindler-CFT cannot be approximated by the global-CFT. }
   \label{fig:TwoFGs}
\end{figure}

One may still wonder if the two CFTs are approximately the same in any sense. After all, when $\rho_c$ and $\rr_c$ are comparable then one would expect the CFTs defined on the two surfaces to be related by conformal transformations. To analyse this, let us consider the global-CFT to see when it can be related to the Rindler-CFT by conformal transformations. In what follows we want $\rho$ to be large so we take $\rho \sim \mathcal O(\epsilon^{-1})$ with $\epsilon \ll 1$. In this limit \bref{GlobalToRindler1}-\bref{GlobalToRindler3} become:
\bea
\rr^2 &=& \rho^2( \sin^2 \phi - \sin^2 \tau) + \cos^2 \tau    ~, \label{GlobalToRindlerLargeRho1} \\
\tanh \xr &=& -(1- \f{1}{2 \rho^2}) \f{\cos \phi}{\cos \tau}  ~, \label{GlobalToRindlerLargeRho2} \\
\tanh \tr &=& (1+ \f{1}{2 \rho^2}) \f{\sin \tau}{\sin \phi}  ~. \label{GlobalToRindlerLargeRho3}
\eea
Equation~\bref{GlobalToRindlerLargeRho1} shows us that we have two distinct possibilities. One is when $( \sin^2 \phi - \sin^2 \tau) \sim \mathcal O(1)$ so that we have $\rr \sim \rho \sim \mathcal O(\epsilon^{-1})$ and the other is when $( \sin^2 \phi - \sin^2 \tau) \sim \mathcal O(\epsilon^2)$ so that we have $\rr \sim \mathcal O(1) \ll \rho$. We consider these two possibilities in detail.
\vspace{.5cm}
\begin{description}
\item[Large $\rr$:]  For this limit we consider $( \sin^2 \phi - \sin^2 \tau) \sim \mathcal O(1)$. We define $\hat \phi = \phi-\pi/2$ and further consider the limit $\hat \phi,\tau \ll 1$. In this limit it is easy to see that we get:
\bea
\rr &=& \rho (1- \f{1}{2} (\tau^2 + \hat \phi^2)) ~,\\
\tr-\xr &=& (\tau-\hat \phi) + \f{(\tau-  \hat \phi)^3}{6} + \f{(\tau+ \hat \phi)}{2 \rho^2} ~, \\
\tr+\xr &=& (\tau+ \hat \phi) + \f{(\tau + \hat \phi)^3}{6} + \f{(\tau- \hat\phi)}{2 \rho^2} ~. 
\eea
Viewed as a diffeomorphism $\rho \to \rr, \tau \to \tr, \phi \to \xr$ the above is realised as a Brown-Henneaux diffeomorphism \bref{BH1}-\bref{BH3} with $\xi^\pm = -\f{1}{6} (\tau \pm \phi)^3$. Note, one can take the limit $\rr \to \infty$ and $\rho \to  \infty$ together and this means that the Fefferman-Graham expansion in the two radial coordinates are consistent. Writing the metric in the Fefferman-Graham form~\bref{FeffermanGraham}\footnote{As explained in footnote~\ref{fn:changeRadialCood} this involves a redefinition of the radial coordinates for both foliations but to avoid clutter we use the same labels.} and using Brown-Henneaux diffeomorphisms we get:
\be
\f{d \rr^2}{\rr^2} + \rr^2 (-d\tr^2  + d\xr^2) + \h(d\tr^2 + d\xr^2) = \f{d \rho^2}{\rho^2} + \rho^2 (-d\tau^2  + d\phi^2) - \h(d\tau^2 + d\phi^2)
\ee
which shows that the negative Casimir energy vacuum state of the global-CFT appears to be an excited state of the Rindler-CFT. 
\\
\\
\item[Small $\rr$:] 
For this limit we consider $( \sin^2 \phi - \sin^2 \tau) \lesssim \mathcal O(\epsilon^2)$. This regions is shown on the boundary cylinder in figure~\ref{fig:TwoFGs}. To show that the Rindler-CFT and global-CFT are not conformally related it suffices to show it in any one part of this region. We consider $\phi \sim \mathcal O(\epsilon)$ and $\tau \sim \mathcal O(\epsilon^2)$. The relations~\bref{GlobalToRindler1}-\bref{GlobalToRindler3} become:
\bea
\rr^2 &=& \rho^2 {\phi}^2+1  \sim \mathcal O(1)~, \\
e^{2 \xr} &=& \f{1}{4} \left[ \f{1}{\rho^2} + {\phi}^2 \right] \sim \mathcal O(\epsilon^2) ~, \\
\tr &=& \f{\tau}{ \phi} (1+ \f{1}{2 \rho^2} + \f{1}{6} {\phi}^2)   \sim \mathcal O(\epsilon)~.
\eea
Similarly the relations~\bref{RindlerToGlobal1}-\bref{RindlerToGlobal3} become:
\bea
\rho &=& \f{e^{-\xr}}{2} \rr \sim \mathcal O(\epsilon^{-1}) ~, \\
\phi &=& 2 \f{\sqrt{\rr^2-1}}{\rr} e^{\xr} \sim \mathcal O(\epsilon) ~, \\
\tau &=& 2  t \f{\sqrt{\rr^2-1}}{\rr} e^{\xr} \sim \mathcal O(\epsilon^2) ~.
\eea
This does not have an interpretation as a small diffeomorphism and hence cannot be interpreted as a Brown-Henneaux diffeomorphism. In addition the coordinate $\rr$ is now $\mathcal O(1)$ so cannot be used as an expansion parameter for Brown-Henneaux diffeomorphisms or Fefferman-Graham expansions. Thus, now one cannot simultaneously take the limit $\rho \to \infty$ and $\rr \to \infty$. 
\end{description}
\vspace{.5cm}
\noindent In summary, the two CFTs are approximated by each other when $\rr$ scales as $\rho$ and this is in the vicinity of the centre of the causal diamond. On the other hand the two CFTs are distinct when $\rho$ is large but $\rr$ is small and this happens in the vicinity of the edges of the causal diamond. The width of the region in which $\rr$ goes from $\mathcal O(\rho)$ to $\mathcal O(1)$ is proportional to the UV cutoff scale of the CFT.

\subsection{Global vs. Poincar\'e boundary} 

\subsubsection{Global and Poincar\'e coordinates and boundaries}

\begin{figure}[htbp] %  figure placement: here, top, bottom, or page
   \centering
   \subfigure[Finite cutoff surfaces]{
   \includegraphics[width=2.2in]{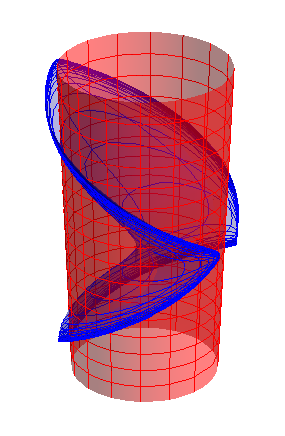}} 
   \hspace{1in}
   \subfigure[Infinite cutoff surfaces]{
   \includegraphics[width=2.1in]{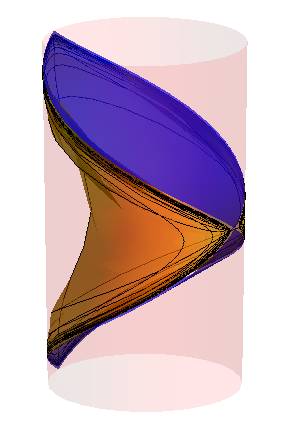}}
    \caption{(a) Global cutoff surface $\rho=\rho_c$ is shown in red and Poincar\'e cutoff surface $\rp=\rp_c$ is shown in blue. (b) When we take $\rho_c$ to infinity, all the $\rp$ surfaces bunch up along the edges of the causal diamond. Two of them are shown in the figure.}
   \label{fig:PoincareCutoffSurfaces}
\end{figure}

The story is analogous for the global vs. Poincar\'e foliations. The metric in the Poincar\'e coordinates is
\be
ds^2 = \f{d\rp^2}{\rp^2} + \rp^2( -d \tp^2 + d \xp^2) \label{PoincareAdS}
\ee
where $\rp \in (0,\infty)$ and $\tp,\xp \in (-\infty,\infty)$. There is a Cauchy (``Poincar\'e") horizon at $\rp=0$. The relation between the global and Poincar\'e coordinates are
\bea
\rho &=& \h \rp \sqrt{ [\rp^{-2}  + (-1+ \xp^2 - \tp^2) ]^2 + 4  \xp^2}   ~,\label{PoincareToGlobal1} \nonumber \\
 \tan \tau  &=& \f{2 \tp}{\rp^{-2}  + (1 +\xp^2 - \tp^2)}~, \label{PoincareToGlobal2} \\
 \tan \phi  &=&- \f{2 \xp}{\rp^{-2} + (-1 +\xp^2 - \tp^2)}~, \label{PoincareToGlobal3}
\eea
and the inverse relations are
\begin{gather}
\rp = \sqrt{1+\rho^2} \cos \tau+\rho \cos \phi~,  \label{GlobalToPoincare1} \\
\tp ~\rp = \sqrt{1 + \rho^2} \sin \tau ~,\label{GlobalToPoincare2} \\
\xp  ~\rp =\rho \sin \phi~. \label{GlobalToPoincare3}
\end{gather}
 The global-CFT is defined on $\rho_c \to \infty$ surface and the Poincar\'e-CFT is defined on $\rp_c \to \infty$ surface. In figure~\ref{fig:PoincareCutoffSurfaces}a we plot two such surfaces without taking the cutoff to infinity.  In figure~\ref{fig:PoincareCutoffSurfaces}b we conformally compactify the global cylinder and see that surfaces of different constant $\rp$ bunch up at the edges of the causal diamond. As in the Rindler-AdS case, imposing boundary conditions on the global boundary imposes conditions on small $\rp$ surfaces also. This justifies our claim that Poincar\'e-CFT  and global-CFT are different.

\subsubsection{Global-CFT vs. Poincar\'e-CFT}

As before we would expect the global and the Poincar\'e-CFT to be approximately the same when $\rho_c$ and $\rp_c$ are comparable since then one would expect the CFTs defined on the two surfaces to be related by conformal transformations. To analyse this we will consider the global-CFT and try to see if and when it can be related to the Poincar\'e-CFT by conformal transformations. We want $\rho$ large so we take $\rho \sim \mathcal O(\epsilon^{-1})$ with $\epsilon \ll 1$. In this limit \bref{GlobalToPoincare1}-\bref{GlobalToPoincare3} become:
\bea
\rp &=& \rho(\cos \tau + \cos \phi) +\cos \tau~,  \label{GlobalToPoincareLargeRho1} \\
\tp  &=&  \frac{\sin \tau }{\cos \tau +\cos \phi } \left(1 +\frac{\cos \phi }{2 \rho ^2 (\cos \tau +\cos \phi
   )} \right) ~,\label{GlobalToPoincareLargeRho2} \\
\xp  &=&   \frac{\sin \phi }{\cos \tau +\cos \phi } \left(1 - \frac{\cos \tau }{2 \rho ^2 (\cos \tau +\cos \phi
   )} \right) ~. \label{GlobalToPoincareLargeRho3}
\eea
Equation~\bref{GlobalToPoincareLargeRho1} shows us that we have two distinct possibilities. The first is when $( \cos \phi + \cos \tau) \sim \mathcal O(1)$ so that we have $\rp \sim \rho \sim \mathcal O(\epsilon^{-1})$ and the other is when $( \cos \phi + \cos \tau) \sim \mathcal O(\epsilon)$ so that we have $\rr \sim \mathcal O(1) \ll \rho$. We consider these two possibilities in detail.
\vspace{.5cm}
\begin{description}
\item[Large $\rp$:]  For this limit we consider $( \cos \phi + \cos \tau) \sim \mathcal O(1)$.  We further consider the limit $\phi,\tau \ll 1$. We get
\bea
\rp &=& 2 \rho (1- \f{1}{4} (\tau^2 + \phi^2)) ~,\\
\tp-\xp &=& \h (\tau-\phi) + \f{(\tau- \phi)^3}{24} + \f{(\tau+\phi)}{8 \rho^2} ~,\\
\tp+\xp &=& \h (\tau+\phi) + \f{(\tau + \phi)^3}{24} + \f{(\tau-\phi)}{8 \rho^2} ~.
\eea
Viewed as a diffeomorphism $\rho \to \h \rp, \tau \to \h \tp, \phi \to \h \xp$ the above is realised as a Brown-Henneaux diffeomorphism \bref{BH1}-\bref{BH3} with $\xi^\pm = -\f{1}{12} (\tau \pm \phi)^3$. One can take the limit $\rp \to \infty$ and $\rho \to  \infty$ together and this means that the Fefferman-Graham expansion in the two radial coordinates are consistent. Writing the metric in the Fefferman-Graham form~\bref{FeffermanGraham}\footnote{As explained in footnote~\ref{fn:changeRadialCood} this involves a redefinition of the global radial coordinate but to avoid clutter we use the same label.} and using Brown-Henneaux diffeomorphisms we get:
\be
\f{d \rp^2}{\rp^2} + \rp^2 (-d\tp^2  + d\xp^2)  = \f{d \rho^2}{\rho^2} + \rho^2 (-d\tau^2  + d\phi^2) - \h(d\tau^2 + d\phi^2)
\ee
which shows that the negative Casimir energy vacuum state of the global-CFT appears to be the vacuum state of the Poincar\'e-CFT with zero energy.
\\
\\
\begin{figure}[htbp] %  figure placement: here, top, bottom, or page
   \centering
   \includegraphics[width=3.5in]{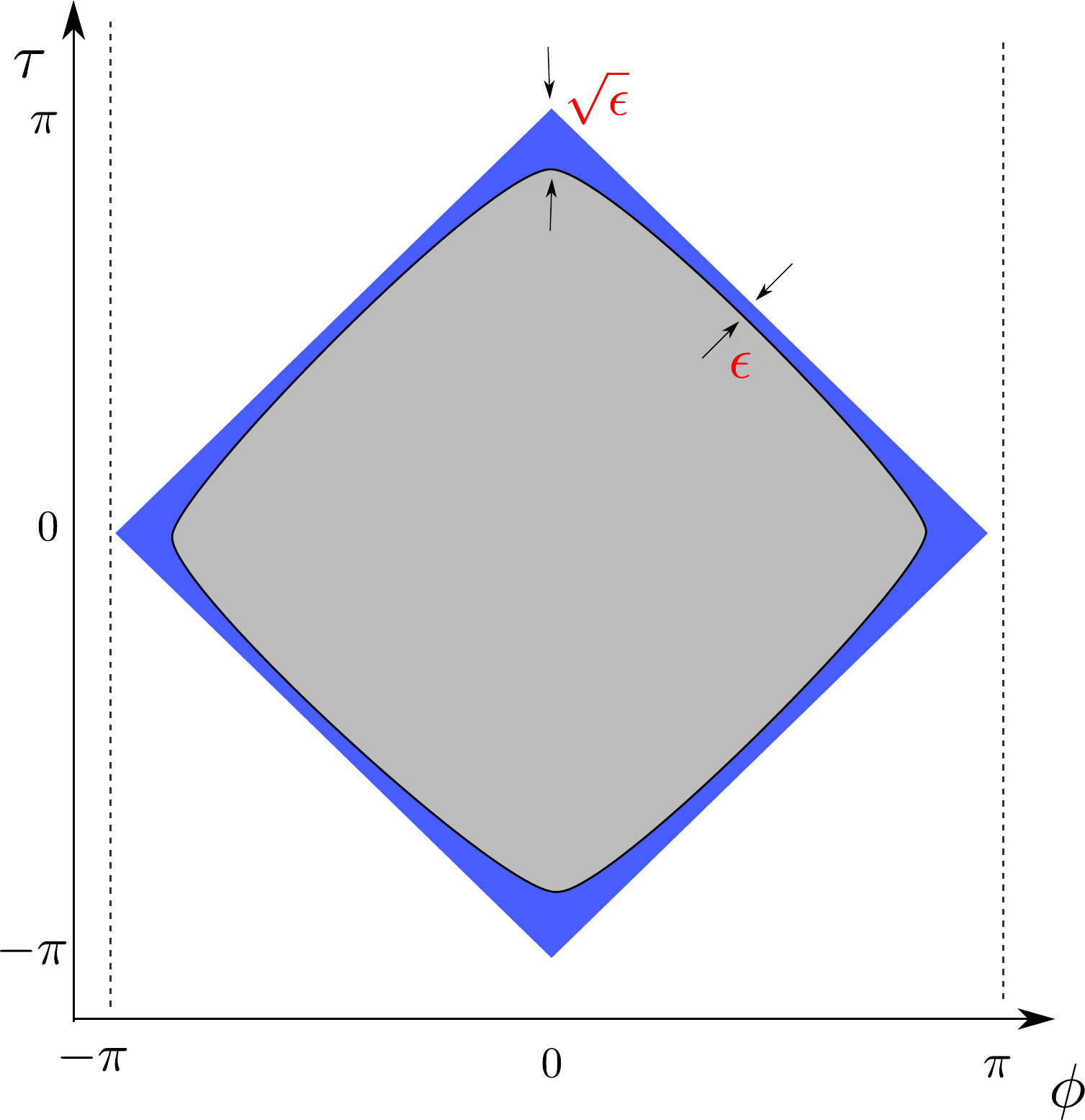}   
    \caption{We open up the boundary cylinder for better visualisation. The coloured region is the causal diamond and is the interior of the curve the bulk Poincar\'e horizon traces on the global boundary cylinder. The boundary cylinder is taken to be large with $\rho \sim \mathcal O(\epsilon^{-1})$ and $\epsilon \ll 1$. The region where the Poincar\'e radial coordinate $\rp \sim \mathcal O(1)$ is shown in blue. This is the region where the Fefferman-Graham expansions in $\rho$ and $\rp$ are not consistent and the Poincar\'e-CFT cannot be approximated by the global-CFT. }
   \label{fig:TwoFGsPoincare}
\end{figure}
\item[Small $\rp$:] 
For this limit we consider $( \cos \phi + \cos \tau) \lesssim \mathcal O(\epsilon)$. This region is shown on the boundary cylinder in figure~\ref{fig:TwoFGsPoincare}. To show that the Poincar\'e-CFT and global-CFT are not conformally related it suffices to show it in any one part of this region. We define $\tilde \phi \equiv \pi - \phi$ and consider $\tilde  \phi \sim \mathcal O(\sqrt{\epsilon})$ and $\tau \sim \mathcal O(\epsilon)$. The relations~\bref{GlobalToPoincare1}-\bref{GlobalToPoincare3} become:
\bea
\rp &=& \h \rho {\tilde \phi}^2+1  \sim \mathcal O(1) ~, \\
 \tp &=& \f{2 \tau}{\tilde \phi^2} \sim \mathcal O(1) ~, \\
\xp &=& \f{2}{\tilde \phi}  \sim \mathcal O(\epsilon^{-1/2}) ~.
\eea
Similarly the relations~\bref{PoincareToGlobal1}-\bref{PoincareToGlobal3} become: 
\bea
\rho &=& \rp \xp^2\sim \mathcal O(\epsilon^{-1}) ~, \\
\tau &=&\f{2 \tp}{\xp^2} \sim \mathcal O(\epsilon) ~, \\
\tilde \phi &=& \f{2}{\xp} \sim \mathcal O(\epsilon^{1/2}) ~.
\eea
This does not have an interpretation as a small diffeomorphism and hence cannot be interpreted as a Brown-Henneaux diffeomorphism. Additionally the coordinate $\rp$ is now $\mathcal O(1)$ so cannot be used as an expansion parameter for Brown-Henneaux diffeomorphisms or Fefferman-Graham expansions. Also now one cannot simultaneously take the limit $\rho \to \infty$ and $\rp \to \infty$. 
\end{description}
\vspace{.5cm}
\noindent As before, we see that the two CFTs are approximated by each other when $\rp$ scales as $\rho$ which is in the vicinity of the centre of the Poincar\'e causal diamond. On the other hand the two CFTs are distinct when the $\rho$ is large but $\rp$ is small and this happens in the vicinity of the edges of the Poincar\'e causal diamond. The width of the region in which $\rp$ goes from $\mathcal O(\rho)$ to $\mathcal O(1)$ is controlled by the UV cutoff scale of the CFT.

\section{Correlation functions in global-CFT, Rindler-CFT, Poincar\'e-CFT} \label{Sec:DifferencesInCFTs}

\subsection{Differences between global-CFT and Rindler-CFT} \label{Sec:DifferencesInCFTsRindlerGlobal}

\begin{figure}[htbp] %  figure placement: here, top, bottom, or page
   \centering
   \subfigure[Rindlerized-global-CFT]{
   \includegraphics[width=2.5in]{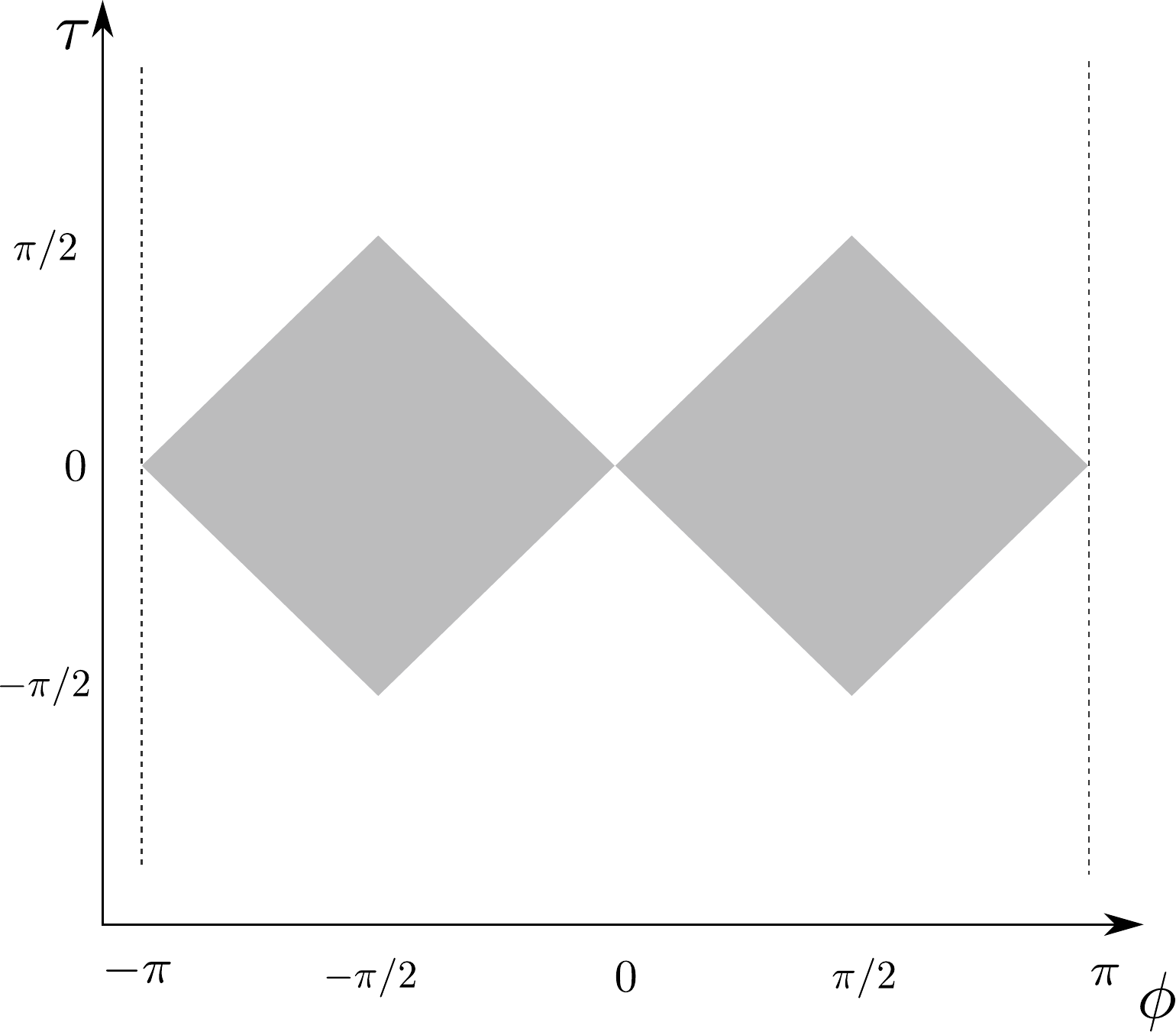}} \hspace{.5in}
    \subfigure[Rindler-CFT]{
   \includegraphics[width=2.5in]{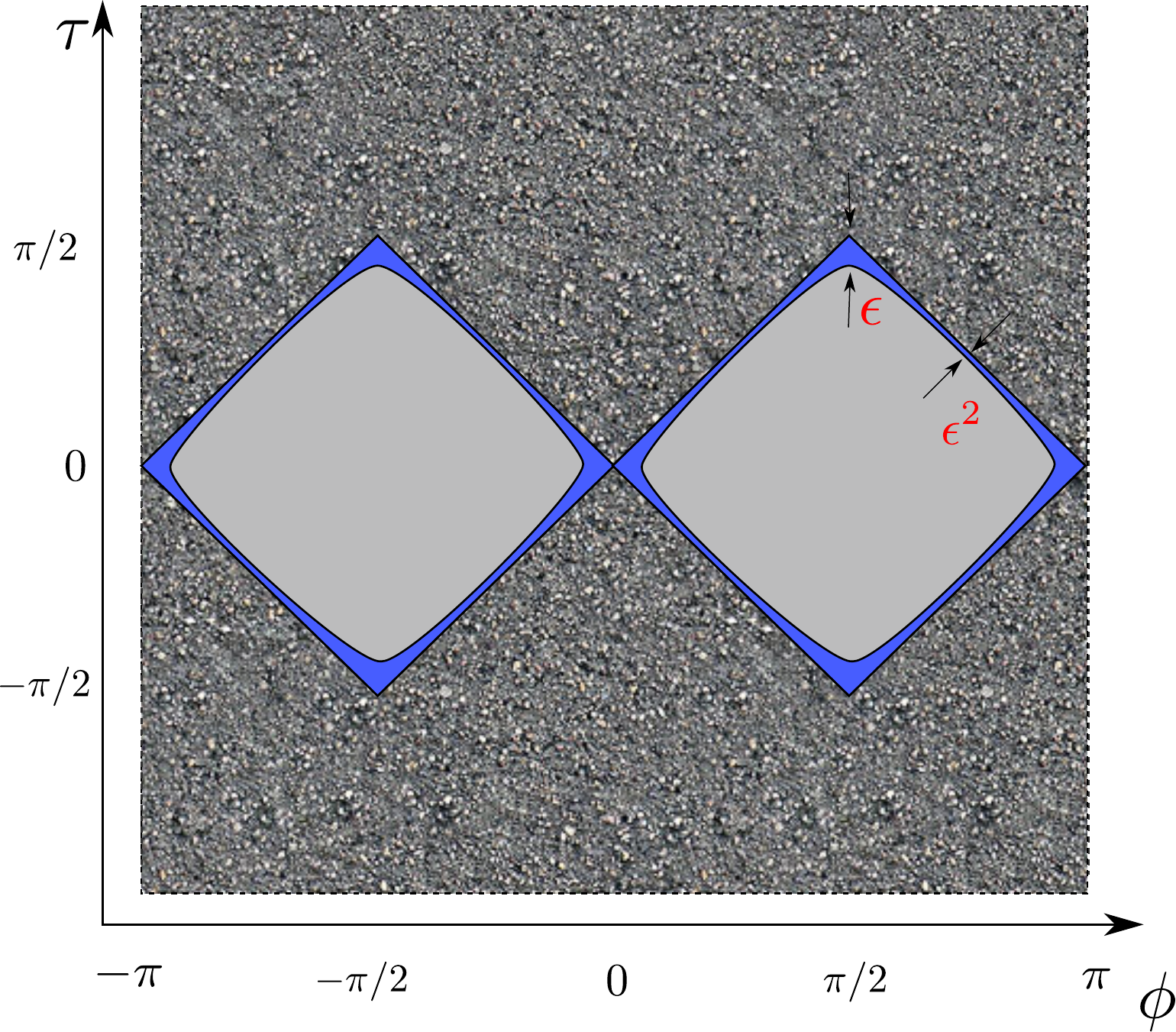}} 
    \caption{We open up the boundary cylinder for better visualisation. In (a) one formally divides the global-CFT into two halves and follows their causal development inside the causal diamonds. This is analogous to ``Rindlerizing the global-CFT''. This is {\em not} the same as the Rindler-CFT (b). While the two theories are approximately equal deep inside the diamonds, they start differing at the edges. The global-CFT is defined everywhere but the Rindler-CFT is defined only  inside the diamonds. Correlation functions of the latter do not give those of the former under analytic continuation.}
   \label{fig:TwoCFTsInDiamonds}
\end{figure}

Consider the coordinate transformations:
\bea
\tanh \xr' &=& -\f{\cos \phi  }{\cos  \tau} ~, \label{Rindlerized-Global-CoodTransformation1} \\
\tanh \tr' &=& \f{\sin \tau  }{\sin \phi} ~.  \label{Rindlerized-Global-CoodTransformation2}
\eea
This is a change of coordinates from the plane to a causal diamond that is the development of $\phi \in (0, \pi)$ in the cylinder. In figure~\ref{fig:TwoCFTsInDiamonds}a we plot the causal diamond (and also its antipodal version).

The two point function of a primary operator $\mathcal O$ of weight $\Delta,\Delta$ on the cylinder is fixed by conformal invariance:
\be
\langle \mathcal O(\tau,\phi) \mathcal O(0,0) \rangle \sim \left( \f{1}{\sin( \f{\tau-\phi}{2})}  \f{1}{\sin( \f{\tau+\phi}{2})}  \right)^{2  \Delta} ~\label{eq:globalCorrelator}
\ee
and under the conformal transformation~\bref{Rindlerized-Global-CoodTransformation1} and \bref{Rindlerized-Global-CoodTransformation2} it becomes
\be
\langle \mathcal O(\tr',\xr') \mathcal O(0,0) \rangle \sim \left( \f{1}{\sinh( \f{\tr'-\xr'}{2})}  \f{1}{ \sinh( \f{\tr'+\xr'}{2})}  \right)^{2 \Delta} ~. \label{eq:twopt-hyp}
\ee
More general correlation functions inside the causal diamonds can be obtained from the ones on the cylinder and conversely the correlation functions inside a causal diamond give the correlation functions on the rest of the cylinder by analytic continuation. 

The coordinate transformations~\bref{Rindlerized-Global-CoodTransformation1} and \bref{Rindlerized-Global-CoodTransformation2} are analogous to the coordinate transformation to go from Minkowski spacetime to Rindler spacetime~\cite{Casini:2011kv} and so, even though the CFT in the diamond is the same as the global-CFT, we refer to it as {\em Rindlerized-global-CFT} as its restricted to the causal diamond.

The new coordinates $\tr',\xr'$ on the global boundary cylinder can be related to the Rindler-AdS coordinates $\tr,\xr$ in the large $\rho$ limit using~\bref{GlobalToRindlerLargeRho2} and~\bref{GlobalToRindlerLargeRho3}:
\bea
\tanh \xr &=& \tanh \xr' (1-\f{1}{2\rho^2})~, \label{RelationRindlerizedGlobalandRindler1} \\
\tanh \tr &=& \tanh \tr' (1+\f{1}{2\rho^2}) \label{RelationRindlerizedGlobalandRindler2}~.
\eea

We can use these to understand what the correlators of the global-CFT imply for the correlators of the Rindler-CFT. In the centre of the causal diamonds $e^{-\xr}, e^{-\tr} \gg \mathcal O(\epsilon)$ and we have $\tr' \approx \tr,~\xr' \approx \xr$. The correlators of the Rindler-CFT can be approximated by
\be
\langle \mathcal O(\tr,\xr) \mathcal O(0,0) \rangle \approx  \left( \f{1}{\sinh( \f{\tr-\xr}{2})}  \f{1}{ \sinh( \f{\tr+\xr}{2})}  \right)^{2 \Delta} ~. \label{eq:approxRindlerCorrelator}
\ee

However, when $e^{-\xr}, e^{-\tr} \sim \mathcal O(\epsilon)$ we get
\bea
e^{-2\tr'} = e^{-2\tr} + \f{1}{4 \rho^2} \\
e^{-2\xr'} = e^{-2\xr} - \f{1}{4 \rho^2}
\eea
and if we are considering the global-CFT then for $e^{-\xr'}, e^{-\tr'} \sim \mathcal O(\epsilon)$ the correlators will continue to be given by~\bref{eq:twopt-hyp} but in the same limit $e^{-\xr}, e^{-\tr} \sim \mathcal O(\epsilon)$ and the Rindler-CFT correlators will now deviate from the form~\bref{eq:approxRindlerCorrelator}.\footnote{To be precise, since we are considering the CFT with a cutoff there will be corrections of the form $\mathcal O((\tr' \pm \xr')/\rho)$ to~\bref{eq:twopt-hyp} and similar corrections to~\bref{eq:approxRindlerCorrelator}. These corrections are of a similar nature as the UV cutoff of both the CFTs are related. However, for the deviations of the Rindler-CFT correlators for large values of $\xr,\tr$ from~\bref{eq:approxRindlerCorrelator} we see that the UV cutoff of the global-CFT induces an IR cutoff for the Rindler-CFT. }

Let us try to understand what this means.   If we were to consider the Rindler-CFT in a thermal state then the correlators would be given by~\bref{eq:approxRindlerCorrelator} for all values of $\tr,\xr$ since the two point function in this case is fixed by conformal invariance. In other words if we were to regulate the Rindler-CFT by putting some boundary conditions at $\xr_{min}=-\tilde x$ and $\xr_{max}=\tilde x$ then the correlators would be different for different boundary conditions but would all approach~\bref{eq:approxRindlerCorrelator} in the limit $\tilde x \to \infty$ irrespective of the boundary conditions.

However, what we see above is that if we consider the global-CFT with a UV cutoff then the Rindler-CFT correlators differ from~\bref{eq:approxRindlerCorrelator} at the edges of the causal diamond. In the limit that the cutoff is pushed to infinity the deviation from~\bref{eq:approxRindlerCorrelator} happens at larger and larger values of $\tr,\xr$ but is always present. 

Conversely, if we consider the Rindler-CFT then the correlators of the Rindlerized-global-CFT will differ from~\bref{eq:twopt-hyp} at the edges of the causal diamond and this immediately implies that the global-CFT correlators will not be the vacuum correlator~\bref{eq:globalCorrelator}. In fact since the Rindler-CFT is not even defined outside the causal diamond it seems likely that analytic continuation of correlators outside the causal diamonds will not work. This is shown in figure~\ref{fig:TwoCFTsInDiamonds}b.

\subsection{Differences between global-CFT and Poincar\'e-CFT} \label{Sec:DifferencesInCFTsPoincareGlobal}

\begin{figure}[htbp] %  figure placement: here, top, bottom, or page
   \centering
   \subfigure[Poincarized-global-CFT]{
   \includegraphics[width=2in]{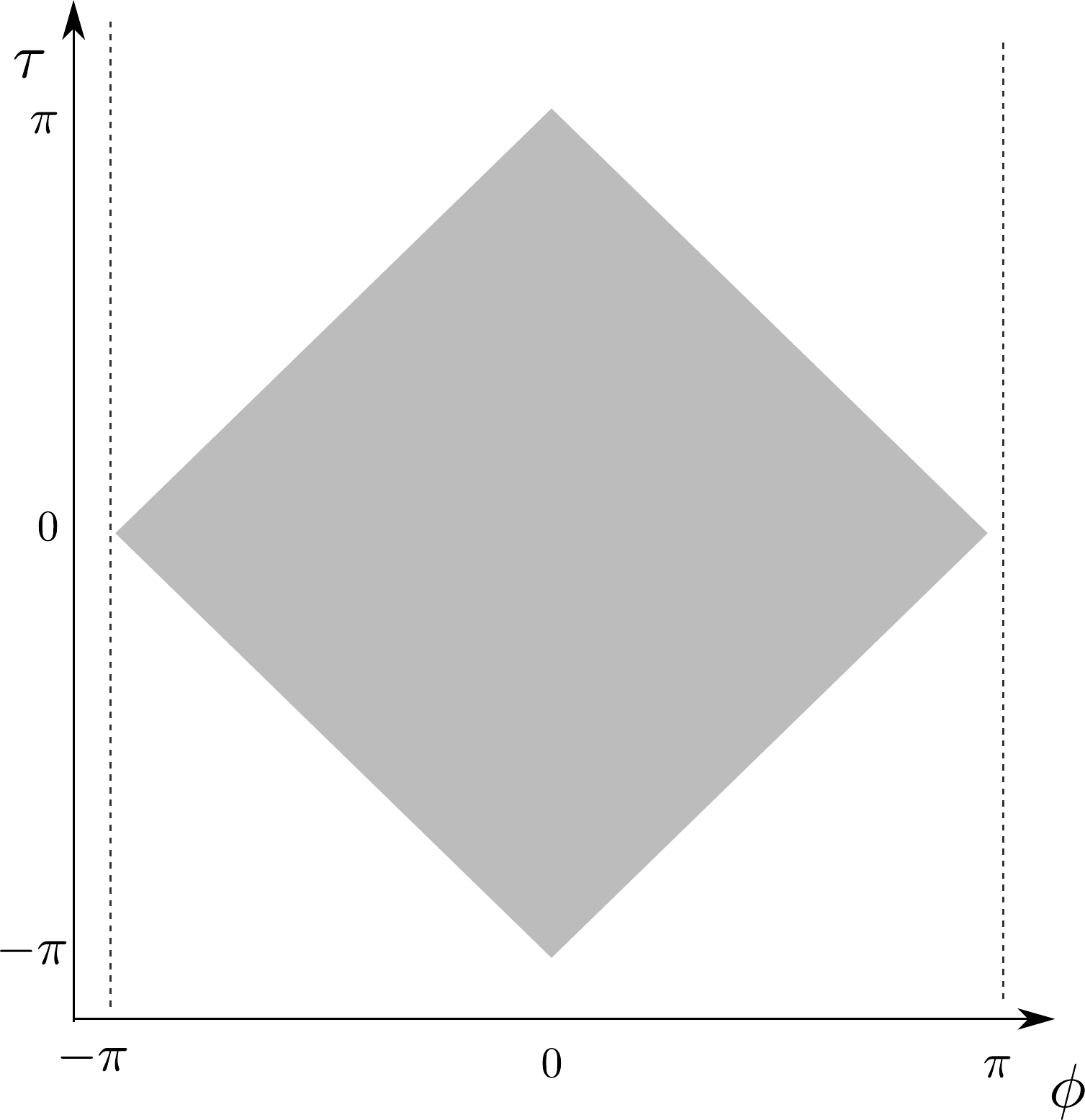}} \hspace{1in}
    \subfigure[Poincare-CFT]{
   \includegraphics[width=2in]{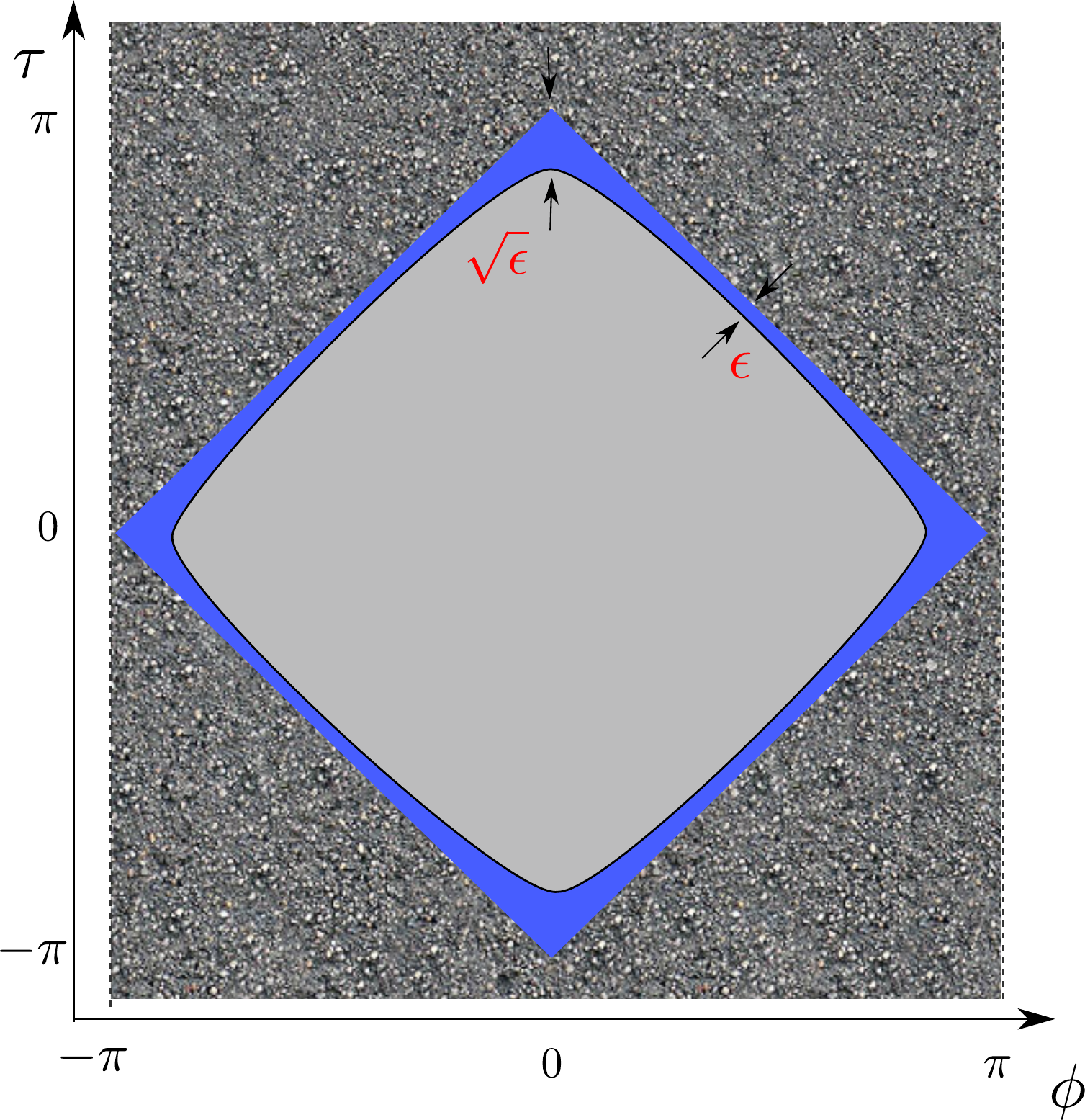}} 
    \caption{We open up the boundary cylinder for better visualisation. In (a) we show the causal development of the interval $(0,2\pi)$. Correlation functions inside can be analytically continued to the full cylinder~\cite{luscher1975}. This is {\em not} the same as the Poincar\'e-CFT (b). While the two theories are approximately equal deep inside the diamond, they start differing at the edges. The global-CFT is defined everywhere but the Poincar\'e-CFT is defined only inside the diamond. Correlation functions of the latter do not give those of the former under analytic continuation.  }
   \label{fig:PoincareDiamond}
\end{figure}

The analysis is similar to that in section~\ref{Sec:DifferencesInCFTsRindlerGlobal} so we will be brief. Consider the coordinate transformations:
\bea
\tp' &=&\f{\sin \tau}{\cos \tau + \cos \phi}~, \label{PoincarizeCood1} \\
\xp' &=& \f{\sin \phi}{\cos \tau + \cos \phi}  ~. \label{PoincarizeCood2}
\eea
This is a change of coordinates from the plane to a causal diamond that is the development of $\phi \in (0,2\pi )$ in the global cylinder. We  open up the cylinder and plot this causal diamond in figure~\ref{fig:PoincareDiamond}a.

 Lucher and Mac~\cite{luscher1975}  showed that correlation functions within this causal diamond can be analytically continued to the whole cylinder. This is because they are the same CFT related by coordinate transformations and so have the same dynamics. We refer to the CFT restricted inside this causal diamond as the {\em Poincarized-global-CFT}. The two-point correlator of a primary operator of weight $\Delta,\Delta$ inside the causal diamond is
\be
\langle \mathcal O(\tp',\xp') \mathcal O(0,0) \rangle \approx  \left( \f{1}{{\tp'}^2 - {\xp'}^2} \right)^{2 \Delta} ~. \label{eq:PoincaredGlobalCorrelator}
\ee

The coordinates $\tp',\xp'$ on the global boundary cylinder can be related to Poincar\'e coordinates $\tp,\xp$ in the large $\rho$ limit using~\bref{GlobalToPoincareLargeRho2} and~\bref{GlobalToPoincareLargeRho3}:
\bea
\tp = \tp' \left (1+ \f{1+ {\tp'}^2-{\xp'}^2}{4 \rho^2} \right ) ~,\\
\xp = \xp' \left (1- \f{1 -{\tp'}^2+{\xp'}^2}{4 \rho^2} \right )  ~.
\eea
We can use these to understand what the correlators of the global-CFT imply for the correlators of the Poincar\'e-CFT. When $\tp,\xp \sim \mathcal O(1)$ we have $\tp' \approx \tp,~\xp' \approx \xp$. This is the center of the causal diamond and the correlators of the Poincar\'e-CFT in this limit have the same expression as those of the Poincarized-global-CFT:
\be
\langle \mathcal O(\tp,\xp) \mathcal O(0,0) \rangle \approx  \left( \f{1}{\tp^2 - \xp^2} \right)^{2 \Delta} ~. \label{eq:approxPoincareCorrelator}
\ee
However, when $\tp, \xp \sim \mathcal O(\epsilon^{-1})$ the two start differing. If we are considering the global-CFT then for $\tp',\xp' \sim \mathcal O(\epsilon^{-1})$ the correlators will continue to be given by~\bref{eq:PoincaredGlobalCorrelator}. But in the same limit $\tp, \xp \sim \mathcal O(\epsilon^{-1})$ and the Poincar\'e-CFT correlators will now deviate from the form~\bref{eq:approxPoincareCorrelator}. 

If we were in the vacuum state of the Poincar\'e-CFT, the correlators would be given by~\bref{eq:approxPoincareCorrelator} for all values of $\tp,\xp$ since this is determined by conformal invariance. In other words if put boundary conditions at $\xp_{min}=-\tilde x$ and $\xp_{max} = \tilde x$ the correlators would be sensitive to these boundary conditions but in the limit $\tilde x \to \infty$ would go to~\bref{eq:approxPoincareCorrelator}. We see precisely this kind of deviation for the Poincar\'e-CFT correlators when considering the global-CFT. Conversely, if we consider the Poincar\'e-CFT then the correlator of Poincarized-global-CFT will differ from~\bref{eq:PoincaredGlobalCorrelator} at the edges of the causal diamond and this implies the global-CFT correlators will be different from~\bref{eq:globalCorrelator}. Since the Poincar\'e-CFT is not even defined outside the causal diamond it seems likely that analytic continuation of correlator outside the causal diamond will not work. This is shown in figure~\ref{fig:PoincareDiamond}b. 

\subsection{Relating the CFTs} \label{sec:RelatingCFTs}

We have argued that the various CFTs are different. An interesting question is whether it may be possible to view the Rindler-CFTs and the Poincar\'e-CFT as deformations of the global-CFT. It seems the answer is yes.

From section~\ref{Sec:DifferencesInCFTsRindlerGlobal} we see that the 
Rindlerized-global-CFTs and Rindler-CFTs are both defined on two copies of $R^{1,1}$. The former is in a particular entangled state (thermofield double state) by construction~\cite{Casini:2011kv} and we can consider the latter in the same state. The two then appear to be the same but there is a subtle difference. In the case of the Rindlerized-global-CFT the Hilbert space associated with the two are the same as the ones associated with the intervals $(0,\pi)$ and $(\pi,2\pi)$ in global coordinates. These two Hilbert spaces are interacting (see section 2 of~\cite{Avery:2013bea}) across the points points $\phi=0$ and $\phi=\pi$. This suggests that one may (roughly) think of the global-CFT as the pair of Rindler-CFTs with an interaction between them across their respective boundaries.  Similarly, one may (roughly) think of the global-CFT as the Poincar\'e -CFT with an interaction across its two ends. 
We hope to make this rough picture more precise in the future. 

\section{Implications for bulk physics} \label{Sec:ConjectureForBulk}

\subsection{A conjecture for the bulk dual of Rindler-CFT and Poincar\'e-CFT}

We have established that correlation functions of Rindler-CFT, Poincar\'e-CFT, and global-CFT are different. Next we would like to understand the implications of this for the bulk physics. In general local bulk physics is quite difficult to examine using the boundary field theory.  However, global causal structures suggest some interesting new physics. 

One might have expected that global AdS is dual to all these CFTs but since their correlators are different, the bulk duals must be different too. Since the correlators are approximately equal till very close to the edges of the causal diamonds and since the causal diamonds are the projections of the event horizons, we conjecture that the bulk duals to the Rindler-CFT and the Poincar\'e-CFT will have a semi-classical description that matches that of the global-CFT till very close to the respective horizons and then start differing. The exact differences will depend on the boundary conditions on the Rindler-CFT and Poincar\'e-CFT; we discuss two specific cases below.

\subsection{Implications for the massless BTZ black hole} \label{sec:MasslessBTZ}

The massless BTZ black hole can be viewed as a quotient of AdS$_3$~\cite{Parsons:2009si} that amounts 
to foliating in Poincar\'e coordinates and periodically identifying~$\xp \sim \xp + 2\pi$. The question we want to ask is {\em what is the bulk dual when we periodically identify the $\xp$ coordinate for the Poincar\'e-CFT?} We refer to this as the periodically identified Poincare-CFT (PIPC).

Let us being by reviewing how the {\em naive} two-point functions of the PIPCs are obtained. Consider Euclidean AdS$_3$ (see appendix~\ref{sec:Euclidean} for details) . The boundary has the topology  $S^2$. According to~\cite{Witten:1998qj}, in Poincar\'e coordinates the boundary is at $\rp=\infty$ (which has a topology $R^2$) with a point at $\rp=0$ added. The boundary to bulk propagator is given by
\be
K_{\text{\tiny P}}^{Euclidean}(\rp,\xp,t_{\text{\tiny E,P}}; \rp^B,\xp^{B}=0,\tp^{B}=0)  \sim \begin{cases}
\rp^{-\Delta} & \text{when} ~\rp^B = 0~, \\
 \left(\f{\rp}{1 + \rp^2[ \xp^2 +\tp^2 ]} \right)^{\Delta}  & \text{when}~ \rp^B = \infty
\end{cases} \label{EuclideanPoincarePropagator}
\ee
where terms with superscripts denote boundary coordinates and $\Delta$ is interpreted as the conformal weight of the associated CFT operator. Naively, one obtains the boundary two-point function for the orbifolded geometry by summing over images of Wick-rotated $\rp^B=\infty$ case of~\bref{EuclideanPoincarePropagator} and then using standard techniques to obtain:
\be
\langle \mathcal O(\tp,\xp) \mathcal O(0,0) \rangle_{\text{\tiny{\cancel{PIPC}}}} \sim \sum_{k=-\infty}^\infty \f{1}{(\xp+\tp+2\pi k)^{2 \Delta} (\xp-\tp+2\pi k)^{2 \Delta}}~. \label{FalseLuninMathur}
\ee

However, there is a problem with this procedure and that is why we have crossed out the subscript on the correlator above. In appendix~\ref{sec:Euclidean} we explain that even in the Euclidean case the asymptotic limits of different foliations are not consistent. In particular on a global $S^2$ of constant radius, large values of $\xp$ correspond to small values of $\rp^B$. In the limit the global radius is taken to infinity, the transition region shrinks to zero in size and is captured by the ``point at $\rp=0$''. For our purposes this implies that the $\rp^B = \infty$ case of~\bref{EuclideanPoincarePropagator} is only correct for $\xp,\tp$ values smaller than the UV cutoff scale. In the Wick-rotated case this means it is valid only inside the Poincar\'e causal diamond away from the edges (see figure~\ref{fig:PoincareDiamond}b). Thus, the method of images can only be used as an approximation deep inside the causal diamond. Since the Poincar\'e-CFT and the global-CFT differ at the edges of the causal diamond and since under periodic identification the edges correspond to late times, the correct two-point function of PIPC would be approximated by~\bref{FalseLuninMathur} for early times but differ at late times. 
Consequently, according to our conjecture the bulk dual should resemble the massless BTZ till very close to the horizon but then should start differing. The length scale over which the transition takes place is governed by the UV cutoff.

Remarkably, we realise in hindsight that in the case of the D1-D5 system which flows in the IR to an $\mathcal N=(4,4)$ CFT all this has been explicitly shown to be the case. Naively the metric and dilaton of the D1-D5 system (with appropriate RR-fluxes) is:
\bea
ds^2_{\text{naive}} &=& \f{1}{\sqrt{g_1 g_5}} ( -d\tp^2 + d\xp^2) + \sqrt{ g_1 g_5} \sum_{i=1}^4 dx^2_i + \sqrt{\f{g_1}{g_5}} \sum_{i=1}^4 dz_i^2 \nn
e^{2 \phi} &=& \f{g_1}{g_5}, \qquad g_1 = 1+ \f{Q_1}{r^2}, \qquad g_5 = 1+ \f{Q_5}{r^2} \label{D1D5Naive}~.
\eea
where the charge radii are related to quantised charges by $Q_{1,5} = g l_s^2 ~ n_{1,5}$. The near-horizon limit of this geometry is massless BTZ $\times T^4 \times S^3$. So when the direction $\xp$ is compactified the D1-D5 CFT plays the role of the PIPC.

However, the actual microstates of the D1-D5 system are not described by the massless BTZ. Instead they correspond to the Lunin-Mathur 2-charge fuzzball geometries~\cite{Lunin:2002qf,Lunin:2002iz,Lunin:2001fv,Skenderis:2006ah,Kanitscheider:2007wq}:
\bea
ds^2_\text{string} &=&  \f{1}{\sqrt{\tilde g_1 \tilde g_5}} ( -(d\tp- A_i dx^i)^2 + (d\xp+ B_i dx^i)^2) + \sqrt{ \tilde g_1 \tilde g_5} \sum_{i=1}^4 dx^2_i \nn & & \ \ + \  \sqrt{\f{\tilde g_1}{\tilde g_5}}  \sum_{i=1}^4 dz^2_i~,  \nn
e^{2 \phi} &=& \f{\tilde g_1}{\tilde g_5}, \quad \tilde g_5 (\vec x) = 1+ \f{Q_5}{L} \int_0^L \f{dv}{|\vec x - \vec F(v)|^2}, \nn   \tilde g_1 (\vec x) &=& 1+ \f{Q_5}{L} \int_0^L \f{ |\dot{ \vec F}(v)|^2 dv}{|\vec x - \vec F(v)|^2}~, \nn
A_i (\vec x) &=& - \f{Q_5}{L} \int_0^L \f{\dot F_i(v) dv}{|\vec x - \vec F(v)|^2}, \qquad d B = - \star_4 ~ d A \label{D1D5metric}~,
\eea
where $\star_4$ is taken with respect to the flat metric for the non-compact $x_i$ space and $v=\xp-\tp$. The solutions are governed by the profile function $\vec F(v)$. The length of integration is given by $L = 2 \pi Q_5$ and one further has:
\be
Q_1 = \f{Q_5}{L} \int_0^L dv (\dot F(v))^2~.
\ee
For these solutions we see that if  $|\vec F(v) | < b$, then at large distances, $r \gg b$, we recover the naive metric \bref{D1D5Naive}. Near $r \lessapprox b$, metrics for different profile functions $\vec F(v)$ differ from each other.

\begin{figure}[ht] %  figure placement: here, top, bottom, or page
\begin{center}
\subfigure[]{\label{fig:blackHoleEmbedding}
	\includegraphics[height=2in]{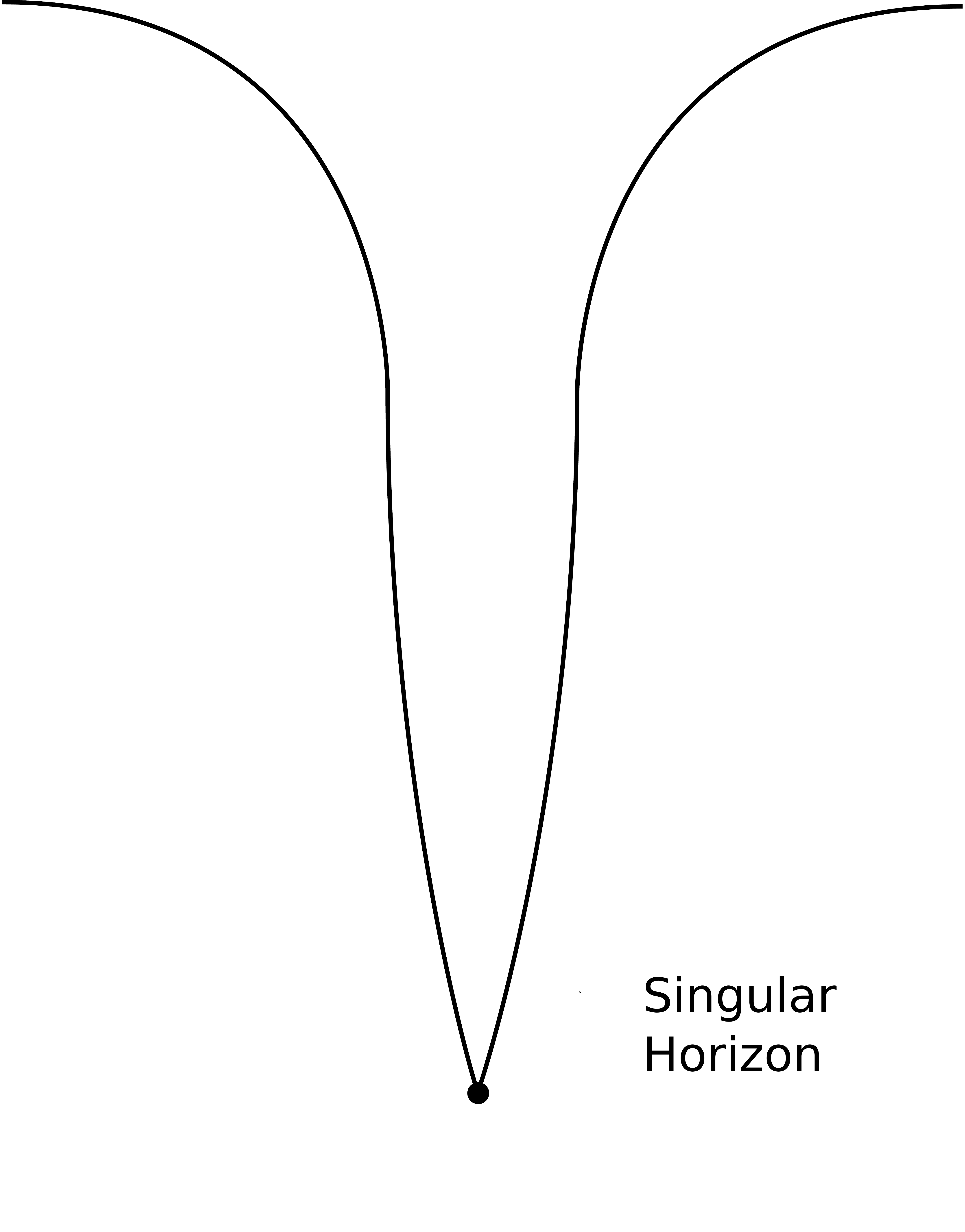}}
\hspace{35pt}
\subfigure[]{\label{fig:fuzzballEmbedding}
	\includegraphics[height=2in]{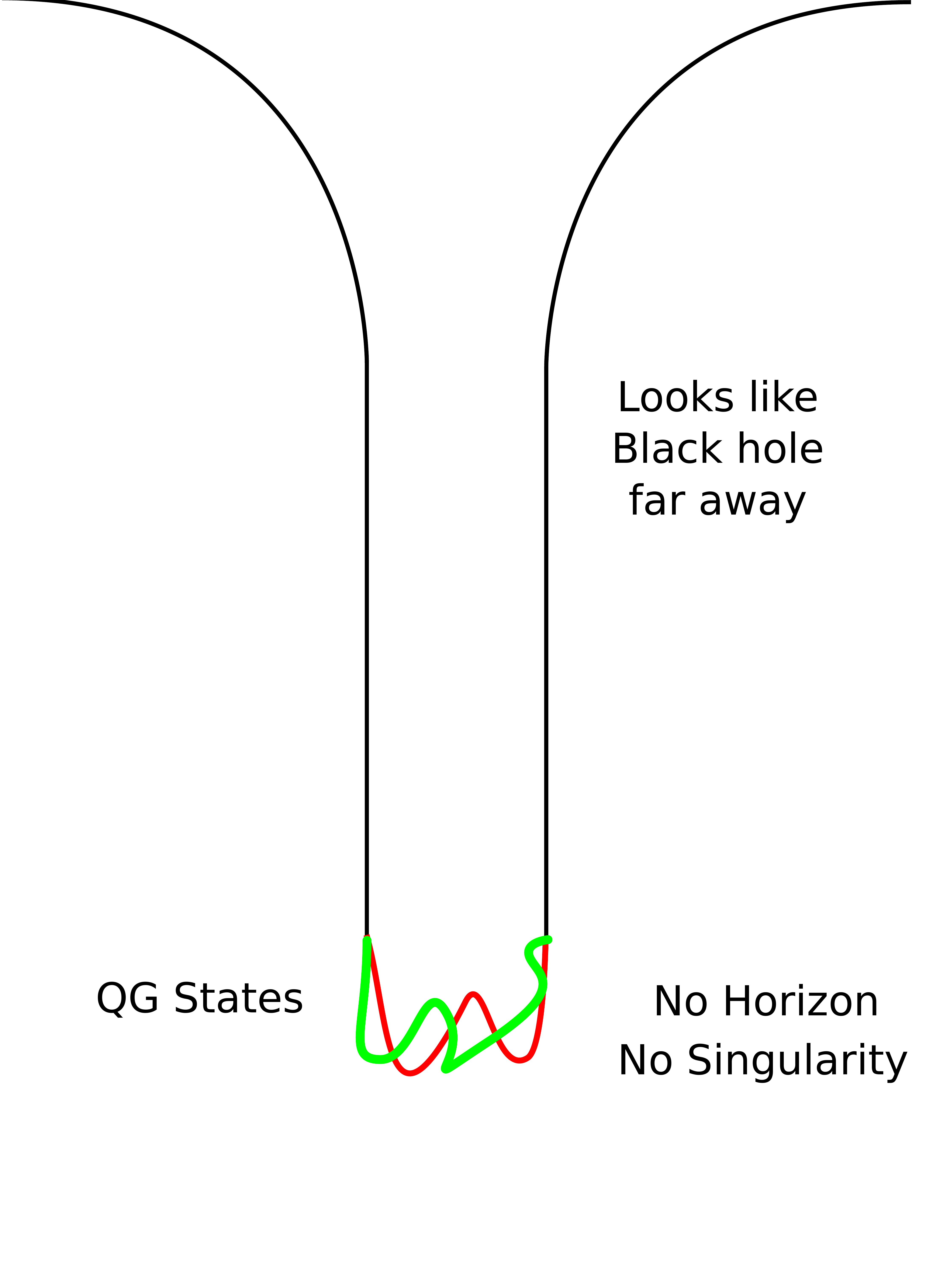}}
   \caption{(a): The geometry of the D1-D5 black hole has an outer flat space connected by a neck to massless BTZ black hole.
   (b) The geometry of a generic state has outer flat space connected by a neck to a throat which ends in a smooth cap without horizons and singularities.}
   \label{fig:BlackHoleNFuzzballEmbedding}
   \end{center}
\end{figure}

The two-point functions in CFT states dual to  these geometries have been worked out in the weak coupling limit~\cite{Balasubramanian:2005qu}. For a subset of these states (conical-defect geometries~\cite{Maldacena:2000dr,Balasubramanian:2000rt}) we reproduce the result:
\be
\langle \mathcal O(\tp,\xp) \mathcal O(0,0) \rangle_{\text{\tiny{conical-defect}}} \sim \sum_{k=0}^{n-1} \f{1}{(2 n \sin \f{\xp +\tp +2 \pi k}{2n})^2 (2 n \sin \f{\xp -\tp +2 \pi k}{2n})^2}
\ee
where $n$ is the order of the conical defect. In general Ref.~\cite{Balasubramanian:2005qu} found from studying the correlators that:
\begin{quote}
``For large central charge (which leads to a good semiclassical limit), and sufficiently small time separation, a typical Ramond ground state of vanishing R-charge has the $M = 0$ BTZ black hole as its effective description.''
\end{quote}
This is consistent with our claim that the true Poincar\'e-CFT correlators agree with the global-CFT correlators deep inside the causal diamond which, after periodic identification, corresponds precisely to {\em sufficiently small time separation}). Ref~\cite{Balasubramanian:2005qu} further states:
\begin{quote}
``At large time separation this effective description breaks down. The CFT correlators we compute take over, and give a response whose details depend on the microstate.''
\end{quote}
This in particular implies that none of the CFT correlators match the naive one~\bref{FalseLuninMathur} at large time separation. This is consistent with our claim that the correlators of the global-CFT and those of Poincar\'e-CFT disagree at the edges of the causal diamond (which after periodic identification just means {\em large time separation}).

Thus in the case of massless BTZ embedded in type IIB supergravity compactified on $S^1 \times T^4$ and $S^1 \times K3$ it is already known that the {\em correct} bulk dual to PIPC is {\em not} obtained from a simple orbifolding of AdS$_3$. Instead the correct bulk duals of PIPC resemble the massless BTZ till very close to the horizon and then quantum gravity effects modify the bulk and cut off the geometry outside the would-be horizons. This is shown in figure~\ref{fig:BlackHoleNFuzzballEmbedding}.

\subsection{Implications for the massive BTZ black hole} \label{sec:MassiveBTZ}

The massive BTZ black hole can be viewed as quotient of AdS$_3$~\cite{Banados:1992gq} that amounts to foliating in Rindler-AdS coordinates and periodically identifying~$\xr \sim \xr +2\pi$. As before, the question we want to ask is {\em what is the bulk dual when we periodically identify the $\xr$ coordinate for the Rindler-CFTs?} We refer to these as the periodically identified Rindler-CFT (PIRCs).

\begin{figure}[htbp] %  figure placement: here, top, bottom, or page
   \centering
   \includegraphics[width=2in]{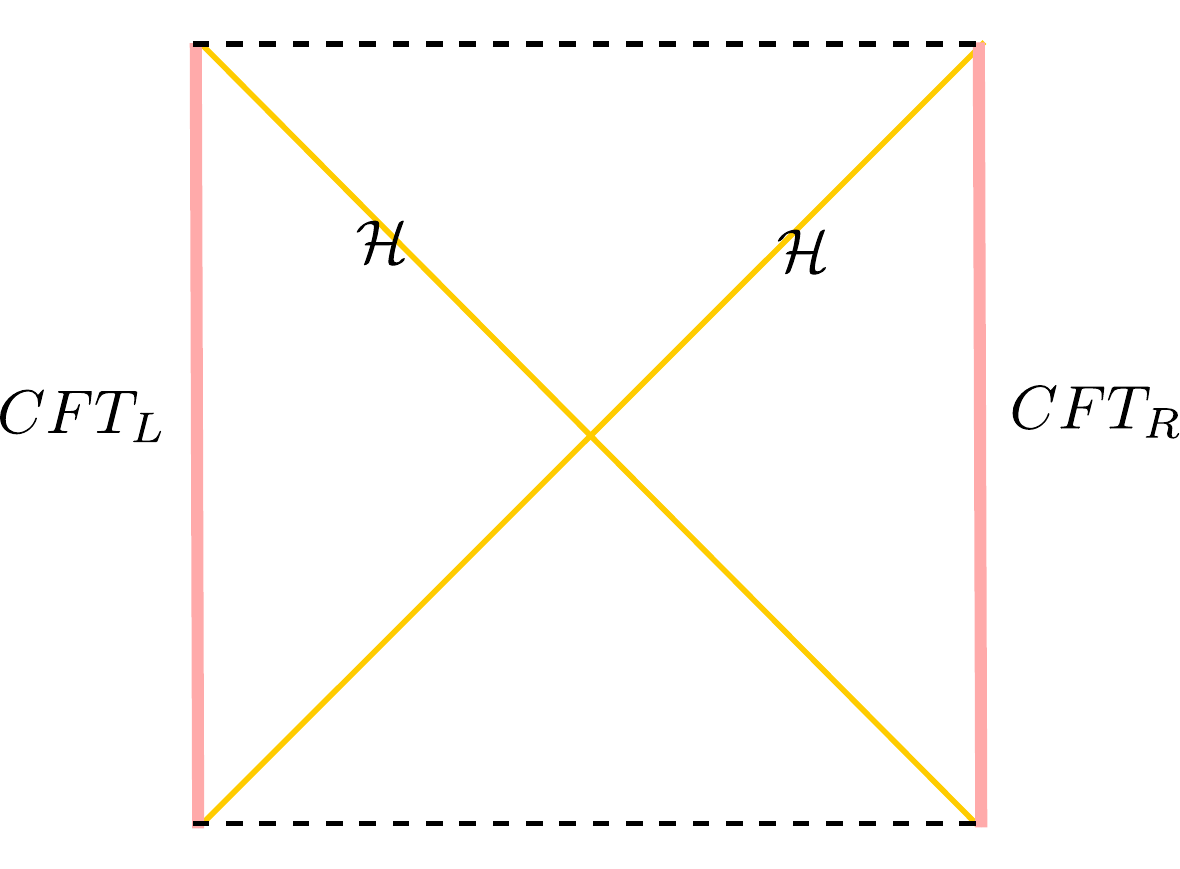}
   \caption{The massive BTZ has two asymptotically AdS region. These regions share the future and past region behind the event horizons.}
   \label{fig:BTZ}
\end{figure}

The Carter-Penrose diagram of the massive BTZ is shown in figure~\ref{fig:BTZ}a.
There are two regions outside the horizons which are asymptotically AdS. These regions have shared future (past) regions behind event horizons which end (begin) in a spacelike singularity. 
When viewed as an orbifold of global AdS$_3$, the event horizons are the Rindler-AdS acceleration horizons and the singularities are orbifolding singularities. 

In~\cite{Maldacena:2001kr}, Maldacena proposed that the massive BTZ black hole is dual to two {\em decoupled} CFTs which can be thought as living on the two boundaries (see figure~\ref{fig:BTZ}) with topologies $R \times S^1$ and are in a particular entangled state called the thermofield double state (for details see~\cite{Maldacena:2001kr}). In particular,  the two CFTs are supposed to capture the dynamics behind the horizons also. This is a remarkable proposal: Imagine two excitations which we call boundary-Alice and boundary-Bob on either CFT. Since the CFTs are decoupled, these excitations have vanishing amplitude to interact. Yet their dual versions, bulk-Bob and bulk-Alice, have a non-zero amplitude to interact in the shared future region. While this may perhaps be possible if we consider a $Z_2$ identification~\cite{Parikh:2002py}, with excitations also respecting the same identification, it seems hard to understand for the usual eternal BTZ black hole. Conceptual puzzles related to this issue have been raised in~\cite{Marolf:2012xe,Avery:2013bea,Mathur:2014dia} and other puzzles with this picture have been raised in~\cite{Kay:2013gia}. 

Nevertheless, one of the reasons to trust this picture has been that orbifolding in the bulk is  innocuous at the horizons, even though its effects are felt inside at the orbifolding singularity. Thus one hopes that the smoothness of horizons would be maintained.  Further, it has been thought that the dual CFT to this orbifolded bulk can be described as quotients of the global-CFT~\cite{Horowitz:1998xk,KeskiVakkuri:1998nw,Maldacena:2001kr,Hemming:2002kd}. One may then hope that the correlators from within the fundamental domain may be  continued to the entire cylinder and this could in principle be a dual description to dynamics behind the horizons.\footnote{We thank Masaki Shigemori for discussions on this point.}

In these kinds of discussions it is either implicitly or explicitly assumed that the boundary CFT for studying Rindler-AdS and that for studying global AdS are the same (see section 2.4 of~\cite{Maldacena:1998bw} and section 2.2 of for example~\cite{Hemming:2002kd} for instance). In this paper we have explicitly shown this {\em not} to be the case. Since the Rindler-AdS radial coordinate $\rr$ becomes the radial coordinate of massive BTZ under orbifolding and the CFT dual to the BTZ is supposed to live on a surface of constant large radius, the CFT associated with the massive BTZ would be the Rindler-CFT instead of the global-CFT. 

This difference is  particularly relevant for the calculation of the CFT two-point function for the PIRCs. The {\em naive} two-point for the periodically identified Rindler-CFT is obtained in the following way. Ref.~\cite{KeskiVakkuri:1998nw} takes the Wick-rotated version of the $\rp^B = \infty$ case of~\bref{EuclideanPoincarePropagator} and uses the large $\rr$ limit of the coordinate transformation between Poincar\'e and Rindler-AdS coordinates:
\bea
\tp \pm \xp &=& \f{\sqrt{\rr^2-1}}{\rr} e^{\tr \pm \xr}~~\to ~~e^{\tr \pm \xr}~,  \\
\rp &=& \rr e^{-\xr}
\eea
to get the bulk-boundary propagator in Rindler-AdS coordinates:
\be
K_{\text{\tiny{\cancel{Rindler-AdS}}}}(\rr,\xr,\tr; \rr^B=\infty,\xr^B=0,\tr^B=0) = \left( \f{ \rr^{-2} e^{- 2 \xr}}{ \left[\rr^{-2} + (1- e^{-\xr -\tr })(1- e^{-\xr + \tr}) \right]^{2}} \right)^{\Delta/2} ~. \label{FalseRindlerAdSPropagator}
\ee
To obtain the boundary two-point function from the orbifolded geometry~\cite{KeskiVakkuri:1998nw} sums over the images of~\bref{FalseRindlerAdSPropagator} and then using standard techniques~\cite{Witten:1998qj} obtains:
\be
  \langle \mathcal O(\xr,\tr) \mathcal O(0,0) \rangle_{\text{\tiny{\cancel{PIRC}}}} \sim  \sum_{n=-\infty}^\infty   \left( \f{1}{\sinh( \f{\tr-\xr-2 n \pi}{2})}  \f{1}{ \sinh( \f{\tr+\xr+2 n \pi}{2})}  \right)^{2 \Delta}  \label{FalseBTZ}~.
\ee
Again, there is a problem with this procedure because of which we have crossed out the subscripts on \bref{FalseRindlerAdSPropagator} and \bref{FalseBTZ}. Not only is the Wick-rotated version of the $\rp^B = \infty$ case of~\bref{EuclideanPoincarePropagator} already an approximation for $\xp,\tp$ deep inside the Poincar\'e causal diamond (as explained in the previous section), but moreover, ~\cite{KeskiVakkuri:1998nw} further takes a large $\rr$ limit which means we are restricting to $\xr,\tr$ deep inside the Rindler-AdS causal diamonds. One cannot then use the method of images on such an approximated propagator~\bref{FalseRindlerAdSPropagator}. Thus, the two-point function~\bref{FalseBTZ} is at best an approximation to the correct propagator for the PIRC for early times. This is good because the correlators~\bref{FalseBTZ} show a large-time decay which is inconsistent with unitary CFTs~\cite{Maldacena:2001kr,Barbon:2004ce,Kabat:2014kfa}.

\begin{figure}[htbp] %  figure placement: here, top, bottom, or page
   \centering
     \includegraphics[width=2.5in]{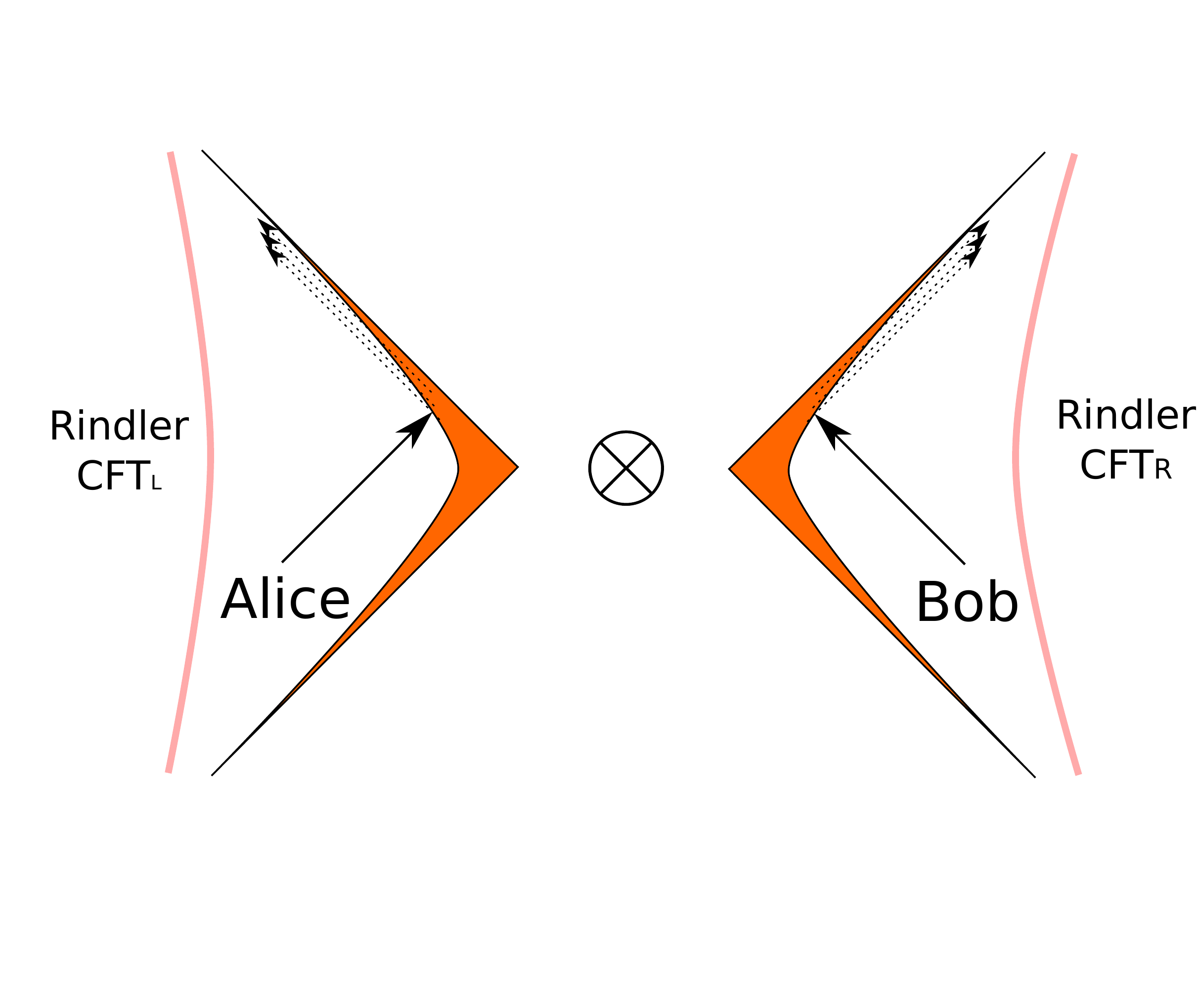}
    \caption{According to the proposals in~\cite{Kay:2013gia,Avery:2013bea,Mathur:2014dia} the bulk dual to two decoupled CFTs on $S^1 \times R$ that are in the thermofield double state (which is equivalent to our PIRCs) are quantum-geometries which resemble the massive BTZ at large distances but get capped outside the ``would-be'' horizons. Thus, there are no shared future or past regions. Our results are consistent with these proposals.}
   \label{fig:conjecture}
\end{figure}

The correct two-point function of PIRC should resemble~\bref{FalseBTZ} till very close to the edge of the causal diamonds. After periodic identification, the edges correspond to late times. So consequently the two should agree at early times but not at late times. Consequently, according to our conjecture the bulk dual should resemble the massive BTZ till very close to the horizon and then start differing. The length scale over which the transition takes place is governed by the UV cutoff.

Our results support the conjecture forwarded by one of us in~\cite{Avery:2013bea} (see also~\cite{Kay:2013gia,Mathur:2014dia}) that states that the true bulk dual to the PIRCs is the one shown in figure~\ref{fig:conjecture} where the bulk dual resembles the massive BTZ outside the horizons but there are no shared future and past regions. The spacetime caps off in ``fuzzballs''.\footnote{For review on fuzzballs see~\cite{Bena:2004de,Mathur:2005zp,Skenderis:2008qn,Balasubramanian:2008da,Chowdhury:2010ct,Bena:2013dka}.} This picture resolves the problems raised in~\cite{Marolf:2012xe,Kay:2013gia,Avery:2013bea,Mathur:2014dia}.\footnote{For an alternate proposal for modifying the bulk see~\cite{Kabat:2014kfa}.}

\section*{Acknowledgements}
It is a pleasure to thank Steve Avery, Jos\'e Barbon, Vijay Balasubramanian, Saugata Chatterjee, Bartek Czech, Jan de Boer, Per Kraus, Gilad Lifschytz, Andrew Long, Gautam Mandal, Juan Maldacena, Samir Mathur, Shiraz Minwalla, Lubo\v{s} Motl, Kyriakos Papadodimas, Joe Polchinski, Kalyana Rama, Suvrat Raju, Eray Sabancilar, Masaki Shigemori, Yogesh Srivastava, Nemani Suryanarayana, Tanmay Vachaspati, Mark Van Raamsdonk and Erik Verlinde for many useful discussions. We would also like to thank Jos\'e Barbon, Vijay Balasubramanian, Bartek Czech, Per Kraus, Gilad Lifschytz, Juan Maldacena, Samir Mathur, Joe Polchinski, Mark Van Raamsdonk and Erik Verlinde for comments on an early draft of this paper. This work was supported in part by DOE grant DE-FG02-09ER41624.
\appendix

\section{Global boundary vs. spherical Rindler boundary}

\begin{figure}[htbp] %  figure placement: here, top, bottom, or page
   \centering
     \subfigure[Finite cutoff surfaces]{
   \includegraphics[width=2.3in]{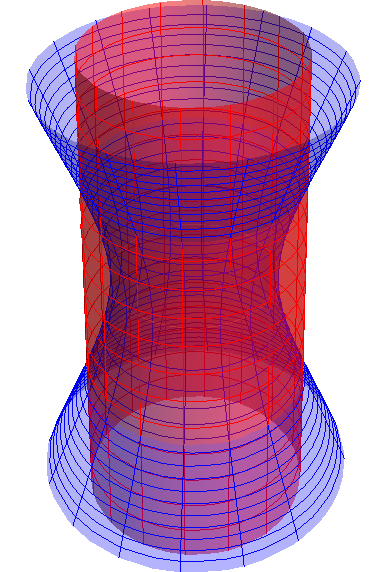}} 
   \hspace{1in}
   \subfigure[Infinite cutoff surfaces]{
   \includegraphics[width=1.6in]{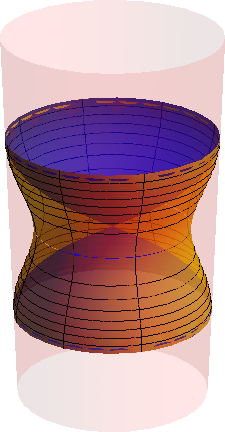}}
    \caption{(a) Global cutoff surface $\rho=\rho_c$ is shown in red and spherical Rindler-AdS cutoff surface $\rsr=\rsr_c$ is shown in blue. (b) When we take $\rho_c$ to infinity, all the $\rsr$ surfaces bunch up along the top and bottom edges of the ``causal strip" which may be thought of as the union of causal diamonds for all the Rindler-AdS observers~\cite{Balasubramanian:2013lsa}. Two such constant $\rsr$ surfaces are shown in the figure.}
   \label{fig:SphericalRindlerCutoffSurfaces}
\end{figure}

In this appendix we study one more foliation of global AdS which involves acceleration horizons in the bulk reaching out to the boundary -- the spherical Rindler-AdS metric.\footnote{Generalisations of this foliation were recently studied in~\cite{Balasubramanian:2013lsa} and similar considerations apply to those foliations also.}  The metric is
\be
ds^2 = \f{d \rsr^2}{\rsr^2-1} +(\rsr^2-1) ~( -d \tsr^2 + \cosh^2 \tsr ~ d\phi^2 )~,
\ee
where $\rsr \in (1,\infty)$, $\tsr \in (-\infty,\infty)$ and $\phi \sim \phi +2 \pi$. The relation between the spherical Rindler-AdS and global coordinates is
\bea
\rho &=& \sqrt{\rsr^2-1} \cosh \tsr ~, \label{SphericalRindlerToGlobal1} \\
\cot \tau &=& \f{\rsr}{\sqrt{\rsr^2-1}} \f{1}{\sinh \tsr}  ~,\label{SphericalRindlerToGlobal2} 
\eea
and the $\phi$ coordinate is the same. The inverse relations are
\bea
\rsr &=& \cosh \rho \cos \tau  ~, \label{GlobalToSphericalRindler1}  \\
\sinh \tsr &=& \f{\cosh \rho \sin \tau}{\sqrt{\cosh^2 \rho \cos^2 \tau -1}} ~.\label{GlobalToSphericalRindler2}
\eea
We see the same theme as before. We can define the global-CFT by fixing boundary conditions on $\rho_c$. However, for large enough $\tsr$, we get $\rsr=1$ surface (and all other small $\rsr$  surfaces) intersecting the global cylinder (see figure~\ref{fig:SphericalRindlerCutoffSurfaces}). Imposing  boundary conditions on the large $\rho_c$ surface puts conditions on small $\rsr$ surfaces also. 
Different boundary conditions suggest that bulk physics would be different also close to the horizon scale. It is possible that the de-Sitter CFT dual to spherical Rindler-AdS may be viewable as a deformation of the global-CFT. We hope to come back to this issue in the future.
 
\section{The Euclidean version} \label{sec:Euclidean}

Euclidean AdS$_{d+1}$ is a ball with the boundary having the topology of $S^d$. Specialising to AdS$_3$ we can write the metric in coordinates which enjoy the symmetries of Euclidean AdS:
\be
ds^2 = \f{d \trho^2}{\trho^2+1} + \trho^2 ( d\ttheta^2 + \sin^2 \ttheta d\tphi^2)~.
\ee
We can also use Poincar\'e coordinates~\bref{PoincareAdS} with Poincar\'e time Wick-rotated:
\be
ds^2 = \f{d\rp^2}{\rp^2} + \rp^2 ( d\tpe^2 + d\xp^2)~.
\ee
One can also foliate $AdS_3$ by Wick-rotating the global time for global coordinate or the Rindler-AdS time for Rindler-AdS coordinates. However, for our purpose the above two suffice.

The relations between these coordinates are
\bea
\sqrt{\trho^2+1} &=& \f{1}{2 \rp} \left( 1+ \rp^2(1+ \xp^2 +\tpe^2) \right)~, \\
\tan \tphi &=& \f{t}{x}, ~ \\
\cos \ttheta &=& \f{\f{1}{2 \rp} \left( 1- \rp^2(1- \xp^2 -\tpe^2) \right)}{ \trho}~, \\
\eea
and the inverse relations are
\bea
\rp &=& \sqrt{\trho^2+1}-\trho \cos \ttheta,~  \label{TildeToPoincare1} \\
 \xp \rp &=& \trho \sin \ttheta \cos \tphi,~ \label{TildeToPoincare2} \\
 \tpe \rp &=&  \trho \sin \ttheta \sin \tphi~. \label{TildeToPoincare3}
\eea
The large $\tilde \rho$ limits of~\bref{TildeToPoincare1}-\bref{TildeToPoincare3} are
\bea
\rp &=& 2 \sin^2(\ttheta/2) \trho,~ \\
 \xp &=& \cot (\theta/2) \cos \tphi \left(1-\f{1}{4 \sin^2(\ttheta/2) \trho^2} \right)~, \\
 \tpe &=&\cot (\theta/2) \sin \tphi  \left(1-\f{1}{4 \sin^2(\ttheta/2) \trho^2} \right)~.
\eea
If we take a large sphere at $\trho \sim \mathcal O(\epsilon^{-1})$ with $\epsilon \ll 1$ then we have two cases. One has $\ttheta \sim \mathcal O(1)$ for which $\rp \sim \trho$ and the other has $\ttheta \sim \mathcal O(\epsilon^{1/2})$ for which $\rp \sim \mathcal O(1)$. In the latter regime $\xp^2 + \tpe^2 \sim \mathcal O(\epsilon^{-1})$. In the limit that the UV cutoff is taken to infinity one can think of the boundary of Euclidean AdS$_3$ being at $\rp=\infty$ with ``a point added at zero''~\cite{Witten:1998qj}. However, as in the Lorentzian case, for any large but finite $\tilde \rho$, surfaces of arbitrarily small $\rp$ intersect the surface of constant $\tilde \rho$.
\bibliographystyle{jhep}
\bibliography{Final}

\end{document}